# Finding faults: A scoping study of fault diagnostics for Industrial Cyber-Physical Systems

Barry Dowdeswell*, Roopak Sinha, Stephen G. MacDonell

*Auckland University of Technology, Auckland, New Zealand*

∗ *Corresponding author. E-mail address: barry.dowdeswell@aut.ac.nz*

**Abstract**

**Context:** As Industrial Cyber–Physical Systems (ICPS) become more connected and widely-distributed, often operating in safety-critical environments, we require innovative approaches to detect and diagnose the faults that occur in them. **Objective:** We profile fault identification and diagnosis techniques employed in the aerospace, auto- motive, and industrial control domains. Each of these sectors has adopted particular methods to meet their differing diagnostic needs. By examining both theoretical presentations as well as case studies from production environments, we present a profile of the current approaches being employed and identify gaps. **Methodology:** A scoping study was used to identify and compare fault detection and diagnosis methodologies that are presented in the current literature. We created categories for the different diagnostic approaches via a pilot study and present an analysis of the trends that emerged. We then compared the maturity of these approaches by adapting and using the NASA Technology Readiness Level (TRL) scale. **Results:** Fault identification and analysis studies from 127 papers published from 2004 to 2019 reveal a wide diversity of promising techniques, both emerging and in-use. These range from traditional Physics-based Models to Data-Driven Artificial Intelligence (AI) and Knowledge-Based approaches. Hybrid techniques that blend aspects of these three broad categories were also encountered. Predictive diagnostics or prognostics featured prominently across all sectors, along with discussions of techniques including Fault trees, Petri nets and Markov approaches. We also profile some of the techniques that have reached the highest Technology Readiness Levels, showing how those methods are being applied in real-world environments beyond the laboratory. **Conclusions:** Our results suggest that the continuing wide use of both Model-Based and Data-Driven AI techniques across all domains, especially when they are used together in hybrid configuration, reflects the complexity of the current ICPS application space. While creating sufficiently-complete models is labor intensive, Model-free AI techniques were evidenced as a viable way of addressing aspects of this challenge, demonstrating the increasing sophistication of current machine learning systems. Connecting ICPS together to share sufficient telemetry to diagnose and manage faults is difficult when the physical environment places demands on ICPS. Despite these challenges, the most mature papers present robust fault diagnosis and analysis techniques which have moved beyond the laboratory and are proving valuable in real-world environments.

**Keywords:** Industrial cyber–physical systems, Faults, Automotive, Aerospace, Avionics, Industrial control

# 1. INTRODUCTION

Industrial Cyber–Physical Systems (ICPS) are mechanisms that augment their computing elements with sensors and electromechanical actuators that allow them to interact with the physical environment they operate in Alur (2015). By evaluating feedback, both from other ICPS they are connected to and from their local industrial environment, they perform a wide range of valuable and often hazardous tasks (Lee et al., 2015). Varying widely in complexity and scale, they are found controlling equipment in aircraft, automobiles and factories.

ICPS should be thought of as being more than just computing devices. They form entire systems, viewed as a collection of seamless entities, including their multiple electrical, mechanical and computing subsystems. This homogeneity makes them fundamentally different to the earlier embedded Programmable Logic Controllers (PLCs) that were first used on General Motors automotive assembly lines in the 1960's (Laughton and Say, 2013; Parr, 1998). These devices controlled only the machinery they were installed or embedded in. They were seldom connected to other plant equipment and the sensors they used were often simpler devices such as limit switches, weight sensors or strain gauges. In contrast, modern ICPS act with higher degrees of autonomy than these earlier embedded systems, relying on sensors and actuators that often incorporate their own local data pro- cessing and conditioning. ICPS are therefore able to make control decisions based on their perception of their environment, driven by much deeper interaction with the physical characteristics of the world they operate in Bajracharya et al. (2008) and Jacoby et al. (2010). Earlier embedded systems seldom featured this degree of complexity and capability.

Contemporary ICPS continue to present intriguing challenges as they have become increasingly more complex. Widely- distributed and now often physically-separated, ICPS are being used to create the Industrial Internet of Things (IIoT), where collections of discrete devices cooperate intelligently to perform large-scale industrial tasks (Leitão et al., 2017). ICPS differ from Cyber–Physical Systems (CPS) used in consumer or medical de- vices primarily in terms of their scale (Yen et al., 2017; Wu et al., 2017), security (Tanveer et al., 2018; Sargolzaei et al., 2016) and safety-critically (Mohrle et al., 2015; Kim et al., 2011). ICPS used in Smart Grids rely on industry-standard interfaces and sophisticated communications. They manage reliable power distribution across wide geographical areas by co-operating and coordinating the operations of the devices that control each sub-station. Examples of advanced ICPS include NASA's Mars rover Curiosity which operates semi-autonomously, controlled by one of the most remote ICPS ever deployed on another planet (Holzmann, 2014; Starek et al., 2016).

Detecting and diagnosing ICPS faults quickly and correctly has become imperative to ensure they are fully-operational at all times. We have learnt how to rely on ICPS more and more to man- age complicated and often safety-critical tasks. Today, undetected failures in ICPS are not just costly: in safety-critical or hazardous conditions they can be life-threatening (McGregor et al., 2017; Feiler et al., 2016). For example, ICPS in the aircraft and aerospace sector rely on accurate readings from sensors to inform guidance, vehicle health and maintain stable flight control. They do this with a degree of reliability, precision and repeatability that human pilots can no longer achieve alone (Chamseddine et al., 2015; Kunst et al., 2009). Similarly, in the automotive sector, vehicles have become increasingly reliant on large local networks of sophisticated subsystems such as anti-skid breaking and fuel- efficient engine controls (Kodali et al., 2013; Sankavaram et al., 2013). Within each subsystem, information is gathered using sensors designed to capture one or more physical characteristics of the local environment, both within and outside the vehicle. The overall operation of a typical ICPS is, therefore, reliant on the co- operative behavior of each of its specialized subsystems, each one dedicated to specific aspects of the vehicle's safe operation and reliability (Schulte, 2018; Shraim et al., 2018; Sankavaram et al., 2016).

### 1.1. The focus and contributions of this study
We identified, categorized and analyzed fault identification and diagnosis strategies for ICPS employed across the aerospace, automotive and industrial control domains. Our goal was to present a snapshot of fault diagnosis as it is practiced today. We surveyed the differences in the approaches that have emerged in each sector and how they address the needs they describe. Our survey provides a guide to applicable techniques for de- signers seeking to implement fault identification, diagnosis and management into their ICPS.

We chose the aerospace, automotive and industrial control domains primarily because the ICPS they rely on must operate faultlessly for extended periods of time, often in close proximity to humans (Bolbot et al., 2018). These sectors also exhibit high levels of integration between their computational cyber elements and the sensors that provide the information that all operational decisions are made on. For example, ICPS in automobiles now sense the position of highway lane markings accurately, extract information from signs and determine the relative positions of adjacent vehicles.

We were also interested in the similarities and differences in fault diagnostic approaches that have emerged in these three safety-critical sectors over the period we studied. The scope of our study was deliberately limited to representative domains that have become highly-dependent on ICPS to manage mission- critical tasks. It is in these sectors that we would expect to find that diagnostics are highly-advanced and widely-used. How- ever, we chose not to include the medical sector in this study. Medical ICPS have distinctive biological characteristics, regulatory requirements and a scale that is worthy of a separate study later. We also excluded cyber–physical devices in the Consumer Electronics sector from our study. They are driving a large and expanding part of the market however they are often less complex than the ICPS in our chosen sectors and the tasks they manage are usually less safety-critical.

A scoping study was used to map the key approaches that underpin fault diagnosis in these sectors and the sources of both theory and case studies available from practitioners (Cacchione, 2016; Arksey and O'Malley, 2005). We framed our study via three research questions:

**RQ1:** What are the most common and widely-used fault identification and diagnosis techniques employed in ICPS in the aerospace, automotive and industrial control domains?

**RQ2:** What relative levels of maturity have the techniques identified in RQ1 achieved when assessed using a systematic scale that is applicable to these domains?

**RQ3:** What research gaps and challenges in ICPS fault identification and diagnosis are being highlighted in the literature surveyed to answer RQ1?

This scoping study seeks to provide a thorough and systematic overview of the fault identification and diagnosis techniques currently in use in our sectors of interest. It profiles the diagnostic approaches we encountered and the techniques that are being used in different situations. By applying a systematic classification to each technique encountered, we are able to estimate the relative level of maturity of each approach, highlighting those which are being applied successfully in real-world environments.

### 1.2. How this paper is organized
Section **2** explores briefly what a ICPS fault is and the terminology used to describe the various stages in a fault management methodology. Section 3 then details the survey data capture and analysis protocol our scoping study employed. While scoping studies do not usually include assessments of the quality of studies uncovered, we chose to adapt and employ the NASA Technology Readiness Level (TRL) as an qualitative scale to compare the relative

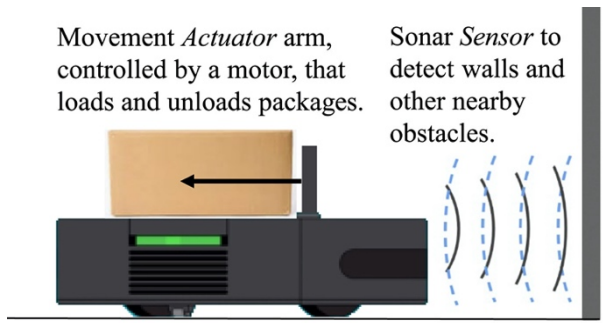

Fig. 1. Sensors and actuators for a warehouse robotic package handler.

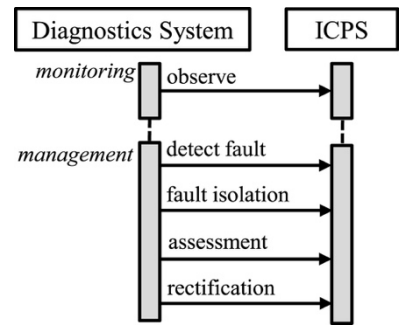

Fig. 2. Sequence diagram of diagnostic activities for a generalized ICPS.

maturity of the fault diagnosis techniques we encountered (Mankins, 2009).

Section **4** presents the results of the scoping study, mapping the fault diagnosis methodologies described in the papers that were included in this study. Finally, Section **5** presents our conclusions, briefly examining those studies that demonstrated the highest TRL. These exemplars discuss fault diagnostic techniques that have moved beyond the laboratory and are being applied in the real world.

## 2. BACKGROUND — WHAT IS FAULT DIAGNOSTICS?

ICPS bridge the connection between their "cyber" software, sensor and actuator hardware parts and the "physical" world they inhabit. Fig. 1 illustrates the two distinct classes of devices that mediate communication across this divide for a warehouse package-handling robot. A sensor is a device that can convert an environmental characteristic such as proximity, pressure, temperature or light levels into an electrical signal that can be pro- cessed by a computer (Jazdi, 2014). In contrast, an actuator is a mechanical device that can receive an electrical signal from a computer and cause a change, often as a result of moving something in its environment (Lee and Seshia, 2016). Motors are special classes of actuators that create movement, such as the mechanism that moves the package off the parcel tray once the robot has arrived at its destination.

Normal behavior for an ICPS such as this warehouse robot is to pick up packages, navigate reliably and efficiently to another location, and then unload its cargo. The robot's activities rely on receiving inputs from its sensors and being able to co-ordinate the movements of its actuators to complete tasks that achieve previously-defined goals. Our example robotic package handler has pre-defined patterns of behavior that enable it to traverse warehouse aisles, locate shelves and deliver packages to specified locations. While it is working, it can detect both obstacles and humans, navigating safely around them.

The difficulty inherent in this interaction between the cyber and physical parts of an ICPS often results in faults occurring. Any change in the way that an ICPS operates that leads to un- acceptable behavior or degraded performance is defined as a fault (Thombare and Dole, 2014). For example, the wheels of the robotic package handler might become entangled with ware- house rubbish from the floor and stop rotating. If the control program detects this problem, it can respond with an appropriate behavior, perhaps stopping and requesting a human for assistance. This sort of situation is not a fault: it is the ICPS managing its behavior in a way that is appropriate. In contrast, not detecting that it cannot move properly and carrying on regardless is a fault since the ICPS did not recognize the issue and change its behavior accordingly. Similarly, failing to detect the edge of stairs and falling down them is unacceptable behavior, possibly due to a faulty precipice sensor. Lee and Seshia comment that it is not enough to separately understand both the computational and electromechanical elements (Lee and Seshia, 2016). Rather, it is at the intersection of the cyber and the physical that the most challenging fault scenarios emerge.

### 2.1. Fault identification, diagnosis and management concepts

Fig. 2 illustrates the activities in a generalized fault management strategy. Fault diagnosis is primarily the analysis of the activities or interactions of an ICPS while it is being observed operating within the environment it is deployed in de Normalisation (2011). Milis et al. (2016) and Harirchi and Ozay (2016) define Fault Detection as the capability of a device to determine the difference between normal and abnormal modes of operation. This may be an after-the-fault examination of a system that has failed or a more proactive monitoring of the system's behavior, watching for issues before they occur. The evidence of a fault is therefore exhibited as unacceptable behavior or degraded performance. Hence, the previous example of the warehouse robot not stopping and requesting assistance is indicative of a fault, either on the part of a sensor or the ICPS software. Fault detection is recognizing that something is wrong but this realization alone does not necessarily categorize or analyze the problem. The purpose of fault detection is to trigger a response by the ICPS to take appropriate action by first recognizing abnormal activity. When faults are detected, the ICPS could just halt. However that is not always a viable strategy if the task the ICPS is performing is critical to some party other than itself.

Detecting faults is the first stage of a Fault Management Strategy (Johnson, 1996). Detecting a fault should start a multi-step process that attempts to diagnose and potentially correct problems so that the ICPS can resume operating at optimal levels. This implies that the ICPS needs to be able to hold a dynamic representation of what normal behavior is so that it can recognize misbehavior.

Fault management strategies include Fault Isolation, which is the process of accurately identifying the location of the fault and its nature (Harirchi and Ozay, 2016). This can be difficult to determine reliably in large systems that contain many interconnected sub-systems. Hence, fault isolation includes the analysis of multiple possible fault sites to determine the nature of the real, underlying fault. The fault symptoms presented, or Fault Evidence, often include secondary system misbehaviors that are the result of the primary fault but which are not the root cause. Bradatsch et al. (2011) defines fault latency as the time between the occurrence of the fault and its recognition by the device's fault management system.

Once a list of possible fault candidate locations has been identified, the next step is Fault Assessment. This examines evidence and seeks to determine the most likely root locations of the fault, as the problem may include a compound failure located at multiple, distinct points (De Kleer and Williams, 1987; Chen, 2011). This leads to the final stage of the diagnosis, Fault Risk Assessment. Not all faults are important enough to require intervention if the system is able to operate satisfactorily in a degraded condition. Ghadhab et al. (2015) discusses the use of "limp-home" strategies for automobiles that allow them to continue to operate safely in a degraded mode until they can be repaired.

Le Mortellec et al. (2013) provide a wider perspective on what an ideal diagnostic system should provide. Besides being able to uniquely identify the true location and nature of a fault, diagnostic systems must be able to communicate effectively with other systems to help facilitate fault rectification. They must deliver their findings rapidly, especially in safety-critical situations. Finally, it is paramount that they must not report false information.

## 3. RESEARCH METHOD

Scoping studies are one method of rapidly mapping the key concepts that appear within a research area (Arksey and O'Malley, 2005; Levac et al., 2010). Often smaller in scope than full systematic reviews or mapping studies, scoping studies allow the breadth of coverage and the depth of the information extracted to be tailored to address research questions appropriately (Wohlin, 2014; Mays et al., 2001). Arksey and O'Malley (2005) and Antman et al. (1992) both explain how scoping studies are an appropriate way to quickly capture and present both the available information and the gaps. They can also be used to focus and inform later literature searches for practitioners when they do not have time to perform a thorough initial analysis themselves. Our scoping study protocol follows the four steps of framing research questions, identifying relevant studies, analysis and then presentation of the results as outlined by Arksey and O'Malley (2005) and refined by Cacchione (2016).

### 3.1. Step one: Framing our research questions
Scoping studies are also effective where the researchers do not have a single or highly-focused research question that they are seeking to answer (Munn et al., 2018). The research questions detailed in Section 1.1 were designed to identify, highlight and categorize practical fault recognition and diagnosis techniques that have been found to be effective both in laboratory studies and in the field. Since this scoping study examines multiple yet similar sectors with potentially differing needs, understanding the focus and spread of the challenges and how they are being addressed should be of interest to practitioners who are designing their own ICPS.

Table 1: Fault identification and diagnosis classification categories.

| Category | Example sub-categories |
|---|---|
| Physics-based modeling approaches | Kalman Filters, Markov Models, Fault Trees, Other Stochiometric processes, Model Validation/Invalidation, Monitor-based oracles |
| Data-driven AI and machine learning approaches | Artificial Neural Networks, Machine Learning, Fuzzy Logic, k-Nearest Neigbour, Big Data/Data Mining |
| Knowledge-based approaches | Bayesian Decision Theory, Binary Trees, Petri nets, Network Message Analysis, Expert Systems |

### 3.2. Step two: Identification of relevant studies
Scopus was used to search for papers that included the terms "cyber–physical", "aerospace", "aircraft", "automotive", "industrial" and "manufacturing" for the fifteen-year period from 2004 to 2019. This starting period for the search was chosen since it coincides with the emergence of the term Cyber–Physical System. The first use of the term can be traced to the National Science Foundation meetings in 2001 that discussed networked embedded control systems (Council, 2001; Gill, 2008). In 2006, Lee (2006) highlighted the implications of these discussions about connecting discrete embedded systems. Prior to this, Wiener's earlier pioneering work on cybernetics informed much of the thinking on control systems theory, arguably setting the agenda for later CPS research (Wiener, 1948).

### 3.3. Step three: Study selection and classification
From an initial pool of 1700 candidate papers returned by our queries, we performed a pilot study on thirty of these pa- pers. Particular papers were chosen primarily because they contained well-written explanations of fault identification and diagnosis techniques that provided valuable background information. These were used to create an initial set of fault identification or diagnosis approach classifications that identified both broad conceptual differences and a list of specific techniques applicable to those approaches. Table 1 lists these categories. RQ1 asks what the nature of fault identification and diagnostics is within our chosen domains. The broadest primary classifications that emerged divided the approaches into three high-level categories that helped to delineate the research activity. We encountered Physics-Based Model-Driven diagnostics, Data-Driven Model-Free Artificial Intelligence (AI) techniques and Knowledge-Based graph approaches. Hybrid techniques that blend aspects of these approaches were also encountered. The similarities and differences between these broad classes are profiled in more detail in Section **4**.

To examine the specific fault-finding methods found within our three primary approaches, sub-categories were created to identify the characteristics of each technique. Beyond these classifications, trends such as Predictive diagnostics or prognostics became of particular interest to us since this approach featured more widely than we initially expected. The complete list of studies, classified according to these category codes, is available via this link (Dowdeswell et al., 2019).

Our publication sources included peer-reviewed journal papers, conference papers and open-access journals. Outside of the academic databases, we also sought technical publications and position papers written by industry-based authors with current, practical experience in their field. Examples include automotive- industry papers from SAE International (https://www.sae.org) and aerospace papers written by NASA researchers or their industry partners (e.g. Lockheed, Boeing). While the papers published by non-academic sources such as SAE were not necessarily peer- reviewed, they often contained detailed results from specific case studies. Arksey and O'Malley stress the importance of including such "gray matter" in scoping studies.

Our minimum inclusion criteria for a study required it to present and explain the fault identification or diagnosis approach that was being applied. We also sought papers that included case studies demonstrating the effectiveness of their techniques. Many papers were excluded because they only mentioned "faults" or "diagnosis" as an aspect of the nature of ICPS without presenting specific examples.

### 3.4. Step four: Analyzing and presenting the data

During the first phase of the analysis, the classification categories allowed us to perform a thematic analysis (Cruzes and Dyba, 2011; Cruzes and Dybå, 2011). Each of our categories and sub-categories represent a technique or approach used or proposed by a practitioner as a way of identifying, diagnosing or rectifying a fault (Castleberry and Nolen, 2018). The analysis also included examining where diagnostic research is focused in each sector and is presented in Figs. **5**, **6** and **7**.

Scoping studies do not usually attempt to assess the quality of the studies uncovered (Cacchione, 2016). However, we chose to adapt and apply a qualitative scale during the classification phase to rank the relative level of maturity of the diagnostic techniques we found. Each study was evaluated using the NASA Technology Readiness Level scale (Straub, 2015; Mankins, 1995). This is a systematic metric for assessing how mature a particular technology is that is now widely used in both aerospace and defense for technology planning. The TRL has been progressively refined since the 1980's through its use at both NASA, ESA and the US military (Mankins, 2009; Straub, 2015; ESA, 2009). It is now embodied in the standard ISO 16290 (ISO/BSS, 2019). In 2014, the European Association of Research & Technology Organisations (EARTO) identified an increased use of the TRL amongst its members as a planning tool to manage innovation (EARTO, 2014).

RQ3 sought to identify research gaps, especially those exhibited amongst the most promising approaches. The TRL provide criteria for assigning a classification between TRL 1, representing basic principles being observed or reported through to TRL 9, characterized by technologies proven in real environments that are ready for widespread adoption. We calibrated our fault diagnostic TRL descriptions using the approach of Terrile et al. (2015). They note that the relative TRL steps are not linear with the steepest steps being in the range TRL 6 to 8. Section **5** details the four divisions we chose to classify studies into an appropriate range. The granularity of the resulting TRL categories allowed us to distinguish between studies that were purely theoretical and those that were profiling fault diagnostic techniques that are being applied in live environments. Table **5** illustrates the fault diagnostic level characterizations we adapted from the NASA categories.

By the end of the classification and analysis phases we had identified fourteen studies that could be ranked at the highest TRLs between 7 and 9. These report mature, field-proven fault- finding and diagnostic strategies that have been deployed in production environments. In those papers, we should expect to see state-of-the-art exemplars that detail how ICPS respond to and recover from fault situations they encounter.

### 3.5. Threats to validity

In scoping surveys such as ours, the primary threats to the validity are our choice of which papers to include and our thematic classifications. Surveys are by definition secondary studies that report broad, summarized characteristics of primary studies, the source papers published that present research about an area of interest (Wohlin et al., 2013). As distinct from Systematic Literature Studies (SLS) that provide highly-detailed evaluations of a smaller set of papers (Kitchenham et al., 2010), scoping studies show where research activity is concentrated and what aspects of a topic are attracting interest, often examining a larger number of papers in less depth.

Internal validity is concerned with the risks that might lead to an incorrect conclusion (Haghighatkhah et al., 2017). This was partially mitigated during the analysis phase by ensuring that each primary paper was initially scanned to determine if it did indeed contain one of our classification classes. For some classes, a list of appropriate synonyms was built iteratively. Our inclusion criteria for a paper included a check to see if groups of related terms were present. The classifications defined in Section **3.3** such as "Model-Based" were expected to show up where models were discussed. However, within the same paper, the classification of "Model-Free" was expected to be applicable when discussions featured AI, Neural Networks, Markov approaches or Data-Driven techniques. Intellectual property restrictions on what can or can- not be published may also be a contributing factor to the amount of detail that can be published about implementations. This was considered when evaluating the relative TRL across sectors.

## 4. DIAGNOSTIC TECHNIQUES IN INDUSTRIAL CPS

Examples of fault identification and diagnostic methods examined initially during the pilot study were described by authors as having evolved along three primary pathways:

Table 2: Physics-based Modeling fault identification and diagnosis techniques across all sectors.

Physics-based Modeling fault identification and diagnosis techniques across all sectors.

| Technique | Aerospace | Automotive | Industrial | All | Publications |
|---|---|---|---|---|---|
| Kalman filters | 23% | 0% | 14% | 13% | Shraim et al. (2018), Goupil et al. (2014), Windmann and Niggemann (2015), Jiang et al. (2018), Chang et al. (2017), Zolghadri et al. (2015), Ruppert and Abonyi (2018) and Ducard (2013) |
| Markov models | 17% | 8% | 9% | 11% | Ghadhab et al. (2015), Windmann and Niggemann (2015), Kunst et al. (2009), White (2012), Ribeiro and Barata (2011), Kodali et al. (2012), Syed et al. (2013), Dearden et al. (2004) and Zhou et al. (2015) |
| Fault trees | 17% | 19% | 5% | 14% | Schulte (2018), Ko et al. (2016), Chen et al. (2014), Hieb et al. (2009), Nuñez and Borsato (2017), Trawczynski et al. (2008), Afonso et al. (2008), Mohrle et al. (2015), Höfig et al. (2015), Ahmad and Hasan (2016), Swearingen and Keller (2007), Klar and Huhn (2012), Syed et al. (2013), Käßmeyer et al. (2016) and Efkemann and Hartmann (2008) |
| Model invalidation | 43% | 41% | 64% | 48% | Provan (2014), Yen et al. (2017), Wang et al. (2017), Palladino et al. (2012), Chen et al. (2012), Goupil et al. (2014), Ko et al. (2016), Chen et al. (2014), Hieb et al. (2009), Nuñez and Borsato (2017), Trawczynski et al. (2008), Afonso et al. (2008), Folkesson et al. (2014), Khlif et al. (2010), Lira and Borsato (2016), Hale and Bollas (2018), White (2012), Barata et al. (2007), Kurien et al. (2002), Luo et al. (2006), Grimm et al. (2012), Schoeller et al. (2007), Ramos et al. (2011), Khlif and Shawky (2008), Castaldi et al. (2016), Harirchi and Ozay (2018), Bartocci et al. (2018), Holzmann (2014), Cossé et al. (2015), Struss (2013), Modest and Thielecke (2016), Khorasgani et al. (2019), Chu et al. (2008), Pakala et al. (2011) and Sanislav et al. (2015) |
| Monitor-based oracles | 10% | 11% | 9% | 10% | Yen et al. (2017), Abel et al. (2013), Schoeller et al. (2007), Kane et al. (2014), Golagha et al. (2017), Dwyer et al. (2010), Ruppert and Abonyi (2018) and Benowitz (2014) |

Physics-Based modeling and analysis frameworks, Data-driven or Model-free AI techniques, and Knowledge-Based graphical approaches (Nuñez and Borsato, 2017). While classifying our studies, we also identified hybrid approaches which blend aspects of these methods.

**4.1. Physics-based, model-driven diagnostics**

Modeling is used by designers to gain a deeper understanding of a system. By creating models that imitate the physical characteristics of the ICPS components, they can explore the interaction the sensors and other physical devices have with the cyber parts of the ICPS (Lee and Seshia, 2016). Physics-Based Modeling techniques for diagnostics rely on consistency checks against these models. These detect the differences between the telemetry captured from the live ICPS and the values predicted by the model. Table 2 summarizes physics-based modeling diagnostic techniques across our survey domains.

Consistency checks use data captured by observers who filter the individual readings to distinguish between noise caused by telemetry errors and values that indicate faulty behavior (Koitz et al., 2017). These differences will often be small but seldom non-zero when the ICPS is performing within acceptable tolerances (Sankavaram et al., 2013). Techniques for determining when an aspect of a model is invalidated were discussed in 48% of papers, especially in the industrial control domain. Both Kalman Filters and Markov Models were discussed as ways of recognizing model invalidation. These techniques implement observers that can process sequential measurements that vary over time. Kalman Filters are more applicable when the range of possible readings is highly-linear. They apply recursive algorithms where weighted-averages are used to estimate the next value. They work well in noisy environments that produce sequences of un- reliable readings. Zolghadri et al. describe an implementation of a Kalman filter to detect jamming of a flight control surface by filtering the error signal before it is processed by the on-board avionics (Zolghadri et al., 2015). The authors explain how the number of sensors providing input to the model affects both the design and worst-case performance. Tuning the model parameters requires trade-offs against the real-time capacities of the diagnostic systems that rely on the model. Shraim et al. discuss fault management for quadrotor unmanned vehicles to improve rotor positioning accuracy (Shraim et al., 2018). Unmanned Aerial Vehicles (UAV) require real-time fault tolerance since they now rely on autonomous, sensor-driven stability control that is no longer managed entirely by the pilot. The models used have to take into account the complex aerodynamic characteristics of the UAV. Dearden et al. discuss similar aspects of autonomous operation, describing fault diagnostics for Mars Rovers where Kalman filtering provides situational awareness to indicate fault conditions (Dearden et al., 2004). They contrast the number of sensors required to manage rover operations with the low computational power available to perform fault identification using multiple sub-system models.

In contrast, Markov models are used to model non-linear, randomly changing systems with discrete states. A dynamic model is Markov or has the Markov Property if the future state of a system depends only on a limited number of previous states. Markov Chain and Markov Decision processes rely on observing the full set of values or states for the aspect of the ICPS that is being diagnosed. In contrast Hidden Markov Models operate where the sequential state of a system is not fully observable. Kunst et al. profile damage propagation through ICPS using Hidden Markov models (Kunst et al., 2009). Similarly, Windmann and Niggemann (2015) and Ribero et al. both apply Markov Models to monitor industrial processes and identify faults as they propagate.

Table 3: Data-driven A.I. Model-Free fault identification and diagnosis techniques across all sectors.

Data-driven A.I. Model-Free fault identification and diagnosis techniques across all sectors.

| Technique | Aerospace | Automotive | Industrial | All | Publications |
|---|---|---|---|---|---|
| Artificial neural networks | 54% | 50% | 47% | 50% | Milis et al. (2016), Sargolzaei et al. (2016), Marzat et al. (2009), Schwabacher and Goebel (2007), Lapira et al. (2013), Sankavaram et al. (2013), Lee et al. (2017), Guo et al. (2016), Yan and Zhou (2018), Langer et al. (2009), Lee et al. (2017), Iverson et al. (2012), Azam et al. (2005), Xu et al. (2012), Banerjee and Das (2013), Schwabacher and Waterman (2008), Vong et al. (2011), Song et al. (2013), Losik (2009), Gonzalez et al. (2009) and von Birgelen et al. (2018) |
| Machine learning | 38% | 17% | 18% | 24% | Wu et al. (2017), Bregon et al. (2014), Sankavaram et al. (2016, 2013), Azam et al. (2005), Xu et al. (2012), Sunny et al. (2018), Fang et al. (2017), Schubert et al. (2011), Losik (2011) and Losik (2009) |
| Fuzzy logic | 8% | 25% | 18% | 17% | Kim et al. (2011), Chen (2011), Khoukhi and Khalid (2015), Song et al. (2010, 2013), Banerjee and Das (2013) and Hu et al. (2011) |
| Big data | 8% | 0% | 18% | 10% | Nagorny et al. (2016) and Schwabacher et al. (2010) |
| Condition monitoring | 8% | 8% | 24% | 14% | Wu et al. (2017), Lee et al. (2015), Fleischmann et al. (2016) and Wang et al. (2017) |

Fault Trees are a way of modeling all reasonably-probable fault scenarios (Schulte, 2018). They are tree structures that facilitate a top-down, systematic approach to identify chains of possible faults. Logical operators can be applied to nodes to identify likely fault pathways. Fault trees are usually considered to be knowledge-based approaches but they were most often encountered in studies that employed hybrid approaches. Mohre et al. demonstrate correlations between fault tree nodes and com- positional safety analysis models (Mohrle et al., 2015). Kassmeyer et al. apply fault trees to track fault scenarios across multiple automotive feature variants (Käßmeyer et al., 2016).

Across all sectors, a wide range of specialized Model-Invalidation approaches were encountered, both theoretical and in- use. Provan (2014) discusses how acceptable inputs can be modeled, an important pre-requisite to detecting misbehavior. Moni- tors (Benowitz, 2014) are code within a fault identification system that is responsible for detecting anomalous situations or behavior. Similarly, Monitor-Based Oracles provide ways of both capturing and evaluating possible fault occurrence (Yen et al., 2017; Abel et al., 2013; Schoeller et al., 2007).

Formal modeling languages including the Architecture Analysis & Design Language (AADL) (Feiler et al., 2006) and Modeling and Analysis of Real-time and Embedded Systems (MARTE) (OMG, 2020) model ICPS during their design phases. AADL originated in the aerospace sector to model embedded systems and has now found wide use in the automotive domain. MARTE extends the UML to provide similar capabilities. Huang et al. (2014) describe a simulation platform modeled in AADL that allows transient faults to be evaluated. Khlif and Shawky demonstrate how to use AADL to design co-simulations that are easier to diagnose later (Khlif and Shawky, 2008). Shulte proposes a state machine architecture for fault detection based on SysML (Schulte, 2018). However, no papers in the survey discussed production ICPS implementations that employed either AADL or MARTE models from the design phases directly. Procter and Feiler present an introduction to the AADL EMV2 Error Library where they discuss the use of an error ontology during modeling (Procter and Feiler, 2020). We searched the literature for examples of the use of EMV2 in production fault diagnostic systems beyond the design phase but found few applicable examples. Lu et al. discuss redundancy approaches using AADL and EMV2 however their work does not demonstrate how to apply their fault trees in a production, real- world example (Lu et al., 2018). Similarly, Zhang et al. discuss the design of fault tolerant systems using EMV2, but it is applicable only to early-stage modeling (Zhang et al., 2017).

Creating and maintaining models is labor-intensive. Many of the techniques rely on detecting situations where a model is invalidated. However, Milis et al. (2016) highlight the amount of effort needed to calibrate models. Provan (2014) also discusses two practical impediments to effective model-based diagnosis: the failure to integrate diagnostic modeling early enough in the requirements process and ambiguities in the models themselves at run-time (see Table 3).

### 4.2. Data-driven fault diagnostics

Data-Driven diagnostic techniques employ training and learning to forge a representation of the system's behavior (Sankavaram et al., 2013). Unlike Physics-Based models, Data- Driven fault detection does not rely on the existence of pre-built models. This approach is preferred when the ICPS can provide telemetry that contains enough information to distinguish between either normal or degraded operations. AI fault diagnosers make sense of that information by using discriminating logic that copes with the changes seen in the ICPS as they occur. This ability to make intelligent decisions distinguishes AI from machine learning, which involves ICPS learning without being explicitly programmed. Milis et al. (2016) discusses cognitive agents that apply expert reasoning to mimic the behavior of human experts.

Artificial Neural Networks (ANN) (Langer et al., 2009; Sargolzaei et al., 2016; Lapira et al., 2013) and pattern-recognition algorithms (Fang et al., 2017) are illustrative of data-driven techniques. Since they do not rely on static, pre-built models as reference points, they remove the need to keep the model up- to-date as the system evolves. Data-Driven diagnostic systems learn behaviors through training. Detection logic allows them to compare current values with previously learnt values (Ramos et al., 2011). Hence these Model-Free methods do not have to completely understand the underlying architecture of the system being examined (Iverson et al., 2012).

Table 4: Knowledge-Based fault identification and diagnosis techniques across all sectors.

| Technique | Aerospace | Automotive | Industrial | All | Publications |
|---|---|---|---|---|---|
| Bayesian networks | 0% | 17% | 43% | 31% | Chen et al. (2012), Chen (2011), Kurz et al. (2011) and Heirung and Mesbah (2019) |
| Binary decision trees | 0% | 33% | 0% | 15% | Waszecki et al. (2015) and Pons et al. (2015) |
| Petri nets | 0% | 0% | 43% | 23% | Yang and Chen (2009) and Cabasino et al. (2010) |
| Network message analysis | 0% | 16% | 29% | 62% | Waszecki et al. (2015), Schweppe et al. (2009), Song et al. (2010) and Sunny et al. (2018) |

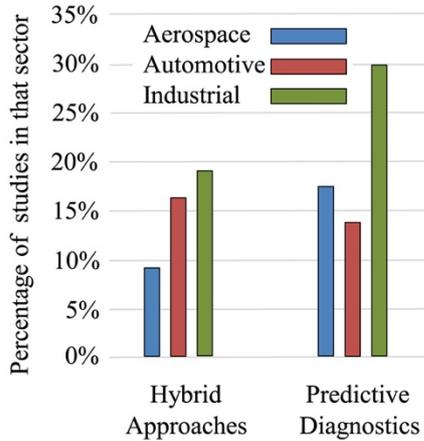

**Fig. 3.** Hybrid and predictive approaches across all sectors.

Data-Driven approaches often scale better than Model-Based techniques (Wu et al., 2017; Yen et al., 2017). As long as sufficient computational resources are available, Data-Driven techniques work as effectively with a large number of sensors as they do with a few (Iverson et al., 2012). Since they construct knowledge representations dynamically, they are often easier to update than formal models (Azam et al., 2005).

Unlike Model-Based methods, Data-Driven approaches do not assume the probabilistic distributions of sampled values that Markov processes rely on (Wu et al., 2017). Similarly, AI methods, including machine learning, do not rely on processes being stochastic or random. The trade-off is that while Physics-Based models are labor-intensive to create, model-free techniques require large example data sets to train the observers (Marzat et al., 2009; Schwabacher and Goebel, 2007; Lapira et al., 2013). Iverson et al. (2012) explain that for avionic ICPS, large volumes of archival sampled values are collected during routine operations that are can be used for training neural networks.

Fuzzy logic employs truth values that are real numbers between zero and one rather than being boolean (Kim et al., 2011; Chen, 2011; Khoukhi and Khalid, 2015). This allows decisions to be made about non-numerical or imprecise data from ICPS, stored in structures called fuzzy sets. These sets represent partial truths and decisions are made by arriving at a consensus. Fuzzy logic algorithms are able to re-evaluate thresholds for situations where values are expected to change dynamically as the system is being observed. Song et al. (2010) discusses recognizing faults using threshold predictions. Each sampled value is checked to see if it falls within a range defined by the previous value read.

Condition monitoring allows Data-Driven fault observers to obtain real-time data about the ICPS they are monitoring. These data points replace the reference values that pre-built Model- Based solutions rely on since AI and machine-learning approaches are model-free (Wu et al., 2017). Lee et al. (2015) and Fleischmann et al. (2016) describe these techniques in terms of system health monitoring. Where deviations from the norm are observed, the result is similar to the model-invalidation discussed earlier. Wang et al. discuss this in the context of cloud computing and predictive maintenance (Wang et al., 2017).

Wang et al. (2018) caution against over-reliance on AI approaches. They suggest that given the complexity of some fault scenarios, the conclusions drawn by data-driven systems may not be sufficiently robust enough to be free of false positives and negatives. However, Iverson et al. profile fault finding for the International Space Station (ISS), reporting that when a large amount of nominal data is available, Data-Driven systems can become highly effective at detecting anomalies (Iverson et al., 2012) (see Table 4).

### 4.3. Knowledge-based approaches

Knowledge-Based approaches are applicable where large amounts of historical data are available. The underlying dependencies that define the system are derived from these sources using a range of techniques. All fault diagnosis systems need to observe real-time data, basing their evaluations on either qualitative or quantitative aspects of the telemetry. However, only knowledge-based approaches utilize significant amounts of historical data to inform their classifiers (Lee et al., 2015). Unlike Data-Driven AI approaches, Knowledge-Based methods do not require pre-classified training sets. Rather, they mine the historical data using statistical methods. Chen et al. explain the value of historical information gathered from experts in building knowledge bases to inform current fault diagnoses (Chen, 2011).

The resultant dynamic models they construct are represented using dependency graphs. Petri nets are directed bipartite graphs where nodes represent discrete fault events that may occur. The graph arcs define possible transitions between states (Yang and Chen, 2009; Cabasino et al., 2010).

Bayesian Belief Networks are knowledge-based directed graphs that model probabilities (Chen, 2011; Kurz et al., 2011). Each node represents a step in a cause and effect chain with a conditional probability. While observing, the fault system up- dates the probability at a node when new information is avail- able. Hence, Bayesian networks can provide both diagnostic and predictive evaluations.

Binary Decision Diagrams are directed acyclic graphs. Waszecki et al. (2015) encode observation patterns extracted from messages exchanged by automotive ECUs to capture fault scenarios that can be evaluated during diagnosis. Network message analysis also complements other knowledge-based approaches, either as a carrier of fault messages or as an indicator of misbehavior (Song et al., 2010). Schweppe et al. (2009) dis- cuss the Automotive Keyword Protocol ISO 14230:2000 (ISO/IEC, 2000), a widely-accepted standard for analyzing faults via net- work messages exchanged over a vehicles CAN bus. Pons et al. (2015) outline a similar approach using Causal Graphs rather than Binary Decision Diagrams.

### 4.4. Hybrid fault diagnostic approaches

Hybrid approaches that blend techniques from any of the three broad approaches were encountered in 14% of the papers but featured in 19% of all industrial control studies. Hybrid techniques skewed the overall ratios of our three primary categories since practitioners can adopt any combination of methods to create their fault identification and diagnostic methodologies. Fig. 3 illustrates the spread of Hybrid approaches across our three domains. Wang et al. (2017) employs Model Invalidation from the Physics-Based Modeling category with Condition Monitoring from the Data-Driven AI category in an intelligent manufacturing scenario. This allows their system to analyze and predict faults from patterns shared via a cloud-based system. The system is implemented using intelligent agents. Chen (2011) combine Bayesian networks with Fuzzy Logic to diagnose faults in auto- motive braking systems while Banerjee and Das (2013) profiles a system with an amalgam of Fuzzy Logic Data-Driven predictors and Model-Based statistical data.

Using multiple approaches in this way allows practitioners to apply the most appropriate technique to different aspects of an ICPS. Rizzoni et al. (2009) discuss how both model-based and neural network techniques facilitated the development of on-board diagnostics and fault monitoring to measure vehicle emissions in automobiles. They trace the motivation for continuously assessing emission compliance in each vehicle back to the California Air Resource Board (CARB) requirements that came into force in 1991. Each vehicle is required to monitor its own emissions to ensure compliance. That required neural network approaches to facilitate the tasks of data capture and sensor filtering followed by model invalidation to test compliance.

### 4.5. Predictive diagnostic techniques

Predictive Diagnostics or Prognostics is the ability to detect the signs of an impending fault before a failure occurs and to estimate when it might happen (Schwabacher and Waterman, 2008). Fig. 3 suggests that the ability to predict ICPS faults in advance is of interest in all three domains. Predictive Diagnostics becomes feasible when it is possible to both capture and process large amounts of high-fidelity data about the operation of an ICPS and recognize the fault symptoms in-advance. Janasak and Beshears (2007) state that one aim of European air carriers is that by 2050, all flights should arrive within one minute of their scheduled time. Current delays and disruptions can be up to fifteen minutes due to undiagnosed faults, an issue that better predictive capabilities might alleviate.

### 4.6. Overall trends in the data

In each sector, there is an emphasis on the development of smart sensors and the conditioning of the sensor data using a range of techniques such as Kalman Filters or Markov models. Coupled with that, the representation of ideal values or behavior was described using either models or dynamically using AI data-mining techniques. Once a definition of what is normal can be determined, deviations from expected values or behaviors can be detected. Artificial Neural Networks and Machine Learning were evidenced as alternatives to Model Invalidation in the Data-Driven AI category. However, the widespread use of hybrid techniques in different parts of the ICPS reflects the complexity of the systems being profiled: no single technique for fault recognition and analysis predominates or is sufficient for all needs. The predominance of Data-Driven techniques in aerospace is in contrast to the lack of evidence for the use of Knowledge- based approaches in that sector while Network Message Analysis was a technique profiled in 29% of the industrial studies that employed Knowledge-based approaches. Those contrasts are explored more deeply in Section 5 where we examine the most mature techniques in more detail.

## 5. INVESTIGATING MATURE FAULT DIAGNOSTIC TECHNIQUES

RQ2 asked what levels of maturity the diagnostic techniques adopted in each sector have achieved. The TRL fault classifications we developed for our study are shown in Table 5 in parallel with the matching NASA descriptions.

Mankins (1995) explains that each level in the TRL scale rep- resents a different maturation of the technology or methodology. Héder (2017) notes that the TRL has drawn criticism for its use outside of the environment it was originally designed for, ex- plaining that in the European Union the approach has not always been tailored properly for specific disciplines. However in NASA the concept of "flight-readiness" was already deeply ingrained in their culture (Feldman, 2000). Adapting this concept to machinery to establish what stage of technological readiness it has reached was a natural step within their context. We considered this when designing our study, carefully crafting our adaptations of the individual level descriptions to ensure we stayed true to the intent of the TRL.

Assessing the maturity of a technological approach requires a careful evaluation of the context that it is being trialled or applied in. Our TRL categories are divided into four distinct groups. Studies classified as TRL 7 to 9 represent the most mature implementations. They provide a fascinating glimpse of techniques which are either close to or fully operational in live production environments.

Studies from TRL 5 and 6 provide evaluations from trials performed in highly-realistic environments beyond the laboratory. They often use case studies to illustrate how the diagnostics will work in particular situations. In contrast, studies at TRL 3 and 4 present functioning prototypes that are being evaluated in either a laboratory or simulated environment.

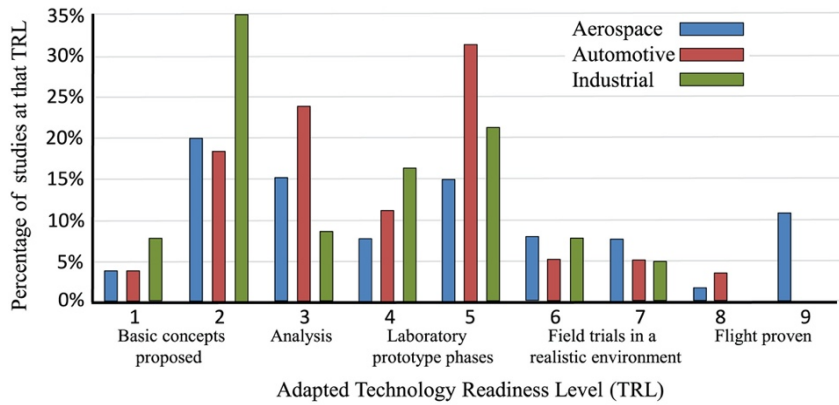

Fig. 4. Technology readiness level by sector.

Table 5: Adapting the NASA Technology Readiness Levels for Assessing Fault Diagnosis.

| TRL | NASA categorization | Proposed fault categorization |
|---|---|---|
| 9 | Actual system "flight proven" through successful mission operations. | Actual fault diagnostic system proven through successful identification and classification of real faults in a production environment. |
| 8 | Actual system completed and "flight qualified" through test and demonstration (ground or space). | Actual fault diagnostic system qualified through test and demonstration in a production environment. |
| 7 | System prototype demonstration in a space environment. | Functioning prototype demonstrated in a production environment. |
| 6 | System/subsystem model or prototype demonstration in a relevant environment (ground or space). | Functioning prototype demonstrated finding and/or diagnosing faults in a relevant environment beyond the laboratory. |
| 5 | Component and/or breadboard validation in a relevant environment. | Creation of a breadboard and/or software validation that can search for and/or identify faults in a relevant environment. |
| 4 | Component and/or breadboard validation in a laboratory environment. | Creation of a breadboard and/or software validation that can search for and/or identify faults in a laboratory environment. |
| 3 | Analytical and experimental critical function and/or characteristic proof-of-concept. | Proof-of-concept experiment with an appropriate simulation of the fault environment. |
| 2 | Technology concept and/or application formulated. | Concept and technology to perform detection and/or diagnosis proposed, including a mathematical formulation. |
| 1 | Basic principles observed and reported. | Basic fault detection or diagnosis principles observed and reported. |

Levels 1 and 2 categorize fault identification and diagnosis techniques that are purely theoretical or are presented with a formal mathematical treatment. Papers at this level do not report concrete outcomes from case studies or field trials.

Fig. 4 highlights the TRL maturity levels we observed across all our domains of interest. Amongst the survey papers are studies of fault identification and diagnostic techniques that have moved beyond the laboratory and are being applied in real-world environments. In these papers, we should expect to see state-of- the-art exemplars that detail how ICPS respond to and recover from fault situations they encounter. The papers we classified at TRL 7 and above present evaluations of how well these techniques detect and analyze faults and why these approaches were adopted.

**5.1. Studies from aerospace and avionics**

Fig. 5 highlights where the diagnostic research is focused in the aerospace sector. The research focus on flight control, high- dependability and predictive fault management aligns with the observations from the studies at the highest TRL discussed in this section.

Benowitz (2014) profiles the Fault Protection Engine currently used by the Mars Curiosity rover. Since the rover is too far away to rely on external systems for assistance, the fault protection engine has to proactively manage faults within a large number of interrelated subsystems autonomously.

Earlier rover designs implemented discrete fault management within each subsystem. On Curiosity, the architecture implements monitors, code within each module whose responsibility is to recognize anomalous behavior. Each module has specific knowledge of the subsystem they are operating within that informs their judgments while filtering sensor readings. Monitors signal problems by raising an error flag. As well as detecting faults, they maintain a count of the occurrences that is later used by the fault protection engine to ascertain how persistent or serious the fault is.

Benowitz explains that error flags are latched but never cleared by the ICPS module-level monitors. This allows the fault protection engine to manage the overall health of the rover by polling in its own time, making decisions without being flooded by messages from subsystems. The fault engine maintains a model that contains a response that is appropriate to each situation the monitors are signaling. Curiosity has over 1000 monitors operating at any one time. Since the rover may be performing any number of different tasks at any time, ranging from landing to exploration, fault management has to be contextual.

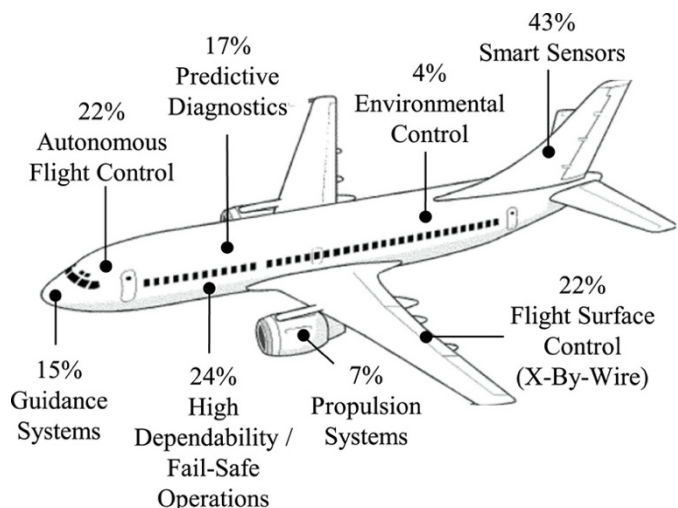

Fig. 5. Where diagnostic research is focused in the aerospace sector.

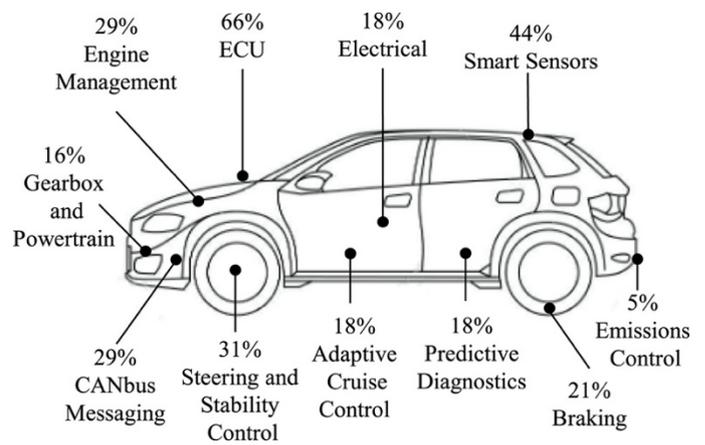

Fig. 6. Where diagnostic research is focused in the automotive sector.

Curiosity's Model-Based approach is in contrast to the hybrid Model-Based and Data-Driven approach employed by Zolghadri et al. (2015). They profile the flight surface control systems they developed for the Airbus A380. Like Curiosity, their fault management is situation-aware. They note that fault signatures are often difficult to detect when an aircraft is parked or taxiing, or when the data rates from sensors are low. Their approach calculates residuals, the result obtained by comparing the current servo positions with the estimated position predicted by the model. They tune the sensitivity of Kalman filters to establish a trade-off between reliably detecting signals and robustness with respect to normal environmental variations. Azam et al. (2005) take a similar approach using neural networks to dynamically model and monitor fifty flight parameters. They discuss the difficulty of using model-based approaches that cannot manage the complexity of accommodating all reasonable parameters in all flight modes. Their data-driven approach also provides estimates of fault severity.

Iverson et al. (2012) and Schwabacher and Goebel (2007), Schwabacher and Waterman (2008), Schwabacher et al. (2009) provide a highly detailed treatment of the hybrid fault monitoring system certified by NASA for International Space Station (ISS) operations and for Ares I-X launch pre-diagnostics. The Inductive Monitoring System (IMS) is a ground-based ICPS that processes telemetry from the ISS in near real-time. It relies on rule-based, Model-Based and Data Driven algorithms in three distinct sub- systems of the IMS. They employ a clustering approach from a fixed number of training points, an approach that allows them to rapidly tailor IMS for new situations. Schwabacher et al. note that there is a need for mission-critical systems such as these to be flight-certified since ground controllers rely on them to make go/no go decisions about launches. They note that many Space Shuttle launches were delayed due to unreliable fault diagnoses. When launch faults can be evaluated more rapidly, redundant or hot-swappable modules can be deployed to reactivate launch sequences to meet critical time windows.

Studies such as these help to explain the proliferation of hybrid techniques encountered. In aerospace, 54% used

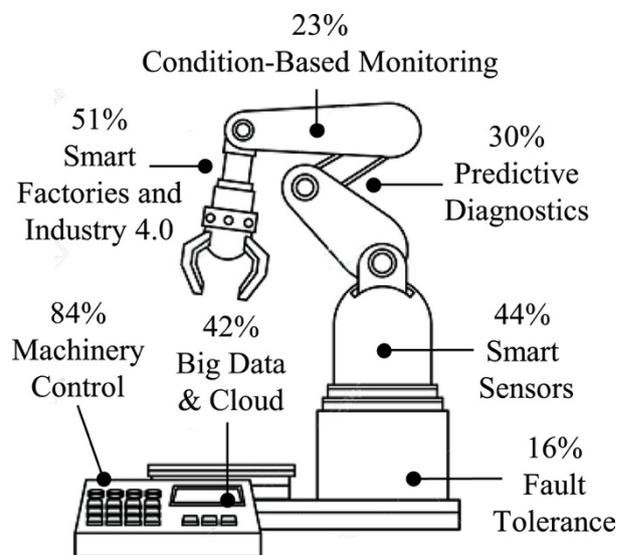

Fig. 7. Where diagnostic research is focused in the industrial control sector.

Artificial Neural Networks and 38% employed Machine Learning, coupled with a range of Model Invalidation methods that were discussed in 43% of all aerospace studies.

### 5.2. Studies from the automotive sector

The automotive ecosystem is built up of millions of discrete, complex and mostly unconnected ICPS. Each vehicle operates as a self-contained network of cooperating subsystems. Stout's Automotive Defect and Recall Report shows that in 2018, nearly eight million vehicles were recalled in the US to address software- based defects (Steinkamp et al., 2019). That total is higher than all the recalls for software issues in the previous five years. Fig. 6 highlights where diagnostic research is focused in the automotive sector.

Modern vehicles feature up to 120 embedded ECUs, connected by five or more system buses (Ebert and Favaro, 2017; Hegde et al., 2011). Sarecco highlights how large and complex the soft- ware is currently in vehicles, reporting that the 2017 Ford 150 pickup requires 150 million lines of code (Saracco, 2016). Charette (2009) contrasts this with the F-35 Joint Strike Fighter that re- quired only 5.7 million lines of code while the Boeing 787 Dream- liner uses only 6.5 million lines.

Table 6: Recurring themes across sectors.

| Theme | Aerospace | Automotive | Industrial Control |
|---|---|---|---|
| Existing degree of connectivity? | Low. | Becoming more connected. | Already highly connected. |
| Difficulty of becoming more connected? | Hard due to distance and low bandwidth. | Hard due to large number of discrete vehicle instances. | Already highly-connected due to high degree of localization. |
| Amount of diagnostic data available? | High. | Becoming very high. | Already very high. |
| Need for autonomous operation? | Very high in remote planetary rovers and drones | Very high due to cost of local fault repair. | Already established via predictive diagnostics and self-management. |

This complexity is reflected in the automotive survey papers at the highest TRL. Nasri et al. (2019) explain that the increasing sophistication of in-car electronics, including Adaptive Cruise Control, Lane Detection and Light Detection And Ranging (LIDAR) technologies, leads to more intricate fault scenarios. They detail the implementation of diagnostics that analyze messages flowing between subsystems on the vehicles Controller Area Network (CAN). Many of the current diagnostic tools rely on proprietary software from vendors that are not easy to integrate into system- wide diagnostic frameworks. They detail their implementation of a hierarchical chain of localized diagnosers that are monitored by a single global fault analyser. A Directed Graph approach is used to identify faults, capturing CAN messages via hardware-in-loop connections.

The scope of what is deemed a "safety-critical" component in the automotive sector is also changing. In May 2018, back-up cameras became mandatory on US vehicles, transforming an optional luxury item into something that required much more rigorous quality control and deeper vehicle integration (Steinkamp et al., 2019).

Over-the-Air (OTA) access to diagnostic data from automobiles is profiled as one route to addressing the difficulty of fault-finding in disconnected automotive ICPS. The global remote diagnostics market is forecasted to grow at 17% annually over the next five years, driven primarily by the potential operational cost savings to automakers (Technavio, 2018). Steinkamp et al. describe General Motors new OTA system which is capable of handling 4.5 TB of data per hour from vehicles (Steinkamp et al., 2019).

However, Dragojević et al. (2018) identify remote access to diagnostic data from a vehicle as a significant technical challenge. Traditional automotive architectures featured highly- specialized ECUs that were optimized for minimal functionality to balance safety concerns. Full operating systems for vehicles emerged though middleware such as Adaptive AUTOSAR (Fürst and Bechter, 2016), leading to greater opportunities to aggregate diagnostic data that could be shared with remote fault analysis systems. Without functionality such as OTA, remote vehicle di- agnostics cannot be performed in an IIoT ecosystem. Dragojevic et al. profile their work on an OTA bridge solution that connects with the on-board vehicle network. However, they note that Adaptive AUTOSAR needs to encompass safety aspects to certifiable levels before it can be widely deployed.

Kane, Fuhrman and Koopman detail the use of runtime monitor-based oracles that mine the data used by OTA systems for fault finding (Kane et al., 2014). Runtime monitors analyze system traces to see if they conform to acceptable behavior patterns. They tune their oracles using large amounts of previously captured telemetry and describe methods used during live vehicle trials. Since monitors operate as hardware-in-loop devices and often interact with safety-critical components, they have to be designed as high-integrity devices. They address this by creating isolated monitors with well-defined interfaces.

### 5.3. Studies from manufacturing and control

Unlike the automotive and aerospace sector, most industrial systems are stationary in one location and are therefore easier to connect into factory-wide monitoring systems. Industrial production machinery therefore offers numerous opportunities to perform local or remote diagnostics. Ramos et al. cite maintenance costs of up to 60% of the production costs as a key driver for factory diagnostics and prognostics (Ramos et al., 2011). Fig. 7 highlights where diagnostic research is focused in the manufacturing and industrial control sector.

International, industry-wide initiatives foster standardization across this sector. Chen et al. (2012) discuss trials of sensors for gearboxes in the context of manufacturing initiatives such as the Machinery Information Management Open Systems Alliance (MIMOSA) (MIMOSA, 2019). Lee et al. (2015) discuss Industry 4.0 and Big Data as similar driver of standardization. Their approach demonstrates end-to-end factory machinery feeding sensor data into multiple analytical systems for near real-time fault identification and prediction. They employ deep-learning for Data-Driven prognostics.

Ramos et al. (2011) also profile Service Orientated Architectures to expose fault-finding services at multiple factory levels. Their case study focuses on self-recovering machinery that is supported by the factory infrastructure using hardware-in-loop techniques. Manufacturing is typically managed by multi-layer IT infrastructures that connect higher-level Enterprise Resource Planning (ERP) through layers down to factory automation systems such a SCADA. Ramos et al. profile their eSonia system which manages assets on multiple levels. Many production operations require assembly lines to be able to be re-configured dynamically to suit changes in demand. This requires a degree of self-awareness from plant equipment,

## 6. CONCLUSIONS, GAPS AND FUTURE WORK

This scoping study was written with a view to providing an overview of mature fault identification and diagnosis techniques for practitioners who are seeking to understand the state of the current practice and who are creating ICPS. The wide use of Model-Based (62%) and alternative Data-Driven AI (33%) techniques across the aerospace, automotive and industrial control domains reflects the complexity of the current ICPS application space.

As the number of interconnected ICPS increases along with the intricacy of the tasks they manage, the use of Model-Based approaches alone was often profiled as becoming intractable. Milis et al. (2016) discussed the difficulties of calibration to align with real systems. Scalability of models was also discussed in this context but only (Yen et al., 2017) discussed partial models, a technique for segmenting models into sub-models. No studies profiled Digital Twins as a solution. Model-Based diagnosis remains a viable strategy, yet how we create complete-enough or partial models quickly and reliably remains a challenge. The AADL EV2 Error Annex has potential to be used beyond the early modeling stages however we found no evidence of its use in the field.

Model-Free AI approaches were evidenced as a viable way of addressing this challenge, demonstrating the increasing sophistication of current machine learning systems. However, there was no discussion of explainable AI, where the decisions made by algorithms could be justified.

The proliferation of hybrid fault systems that blend different aspects and techniques reached 19% in the industrial control sec- tor, indicating the importance of further research into multiple- method solutions, where models are tuned by real-time data. Design-for-Certification was highlighted as a significant driver to ensure products could be deployed beyond the laboratory (Dragojević et al., 2018; Schwabacher and Goebel, 2007; Schwabacher et al., 2010).

Predictive diagnostics is a promising area that was often discussed in-context with the ability to mine sensor data with enough granularity to allow faults to be predicted. Predictive techniques were prevalent in 30% of all industrial control studies, driven by the availability of large amounts of local data. Further research to develop remote connectivity in the aerospace and automotive sectors should lead to more powerful predictive capabilities. However, the potential volume of the data available from these ICPS also presents challenges of scale.

Statistical aspects of Knowledge-based diagnostic approaches were poorly represented across the aerospace sector. Most applications of the technique in the automotive and industrial control sectors discussed Bayesian approaches and various Petri net derivatives. This may be due to the increasing presence of hybrid approaches which employ Knowledge-Based methods in the midst of other techniques. There was little evidence of traditional Expert Systems.

Connectivity is a key characteristic of ICPS yet it has deeper implications in our sectors of interest. Table 6 illustrates how connectivity for facilitating diagnostics is made more challenging because of the different environments ICPS operate in. Brief discussions in the papers of emerging cloud technologies pointed towards ways of establishing connectivity in more achievable ways.

While the TRL analysis provided a way of identifying and pro- filing the most mature approaches, those results cannot always be extrapolated across all three sectors. Almost all the avionic and aerospace studies profiled originated from organizations who were partnering with agencies such as NASA and ESA. These do not face the same intellectual property restrictions that restrict what we might expect to find published in the automotive and industrial control sectors.

During our paper selection, promising papers from the medical device ICPS sector gave a tantalizing glimpse of the differences and challenges that sector presents. We look forward to exploring that domain in a later study, where complex, safety-critical devices and regulatory certification are the norm rather than the exception.

**CRediT authorship contribution statement**
**Barry Dowdeswell:** Conceptualization, Methodology, Data curation, Investigation, Writing - original draft, Writing - review & editing, Project administration. **Roopak Sinha:** Conceptualization, Supervision, Methodology, Writing - review & editing, Validation. **Stephen G. MacDonell:** Conceptualization, Supervision, Methodology.

**Declaration of competing interest**
The authors declare that they have no known competing financial interests or personal relationships that could have appeared to influence the work reported in this paper.

## REFERENCES

Abel, A., Adir, A., Blochwitz, T., Greenberg, L., Salman, T., 2013. Development and verification of complex hybrid systems using synthesizable monitors. In: Haifa Verification Conference. Springer, pp. 182–198. http://dx.doi.org/10.1007/978- 3- 319- 03077- 7_13.

Afonso, F., Silva, C., Tavares, A., Montenegro, S., 2008. Application-level fault tolerance in real-time embedded systems. In: 2008 International Symposium on Industrial Embedded Systems. IEEE, pp. 126–133. http://dx.doi.org/10.1109/SIES.2008.4577690.

Ahmad, W., Hasan, O., 2016. Formalization of fault trees in higher-order logic: a deep embedding approach. In: International Symposium on Dependable Soft- ware Engineering: Theories, Tools, and Applications. Springer, pp. 264–279. http://dx.doi.org/10.1007/2F978- 3- 319- 47677- 3_21.

Alur, R., 2015. Principles of Cyber-Physical Systems. MIT Press.
Antman, E., Lau, J., Kupelnick, B., et al., 1992. A comparison of results of meta- analyses of RCTs and recommendations of clinical experts. Treatments for myocardial infarction. JAMA 268, 240–248.

Arksey, H., O'Malley, L., 2005. Scoping studies: towards a methodological framework. Int. J. Soc. Res. Methodol. 8 (1), 19–32. http://dx.doi.org/10.1080/1364557032000119616.

Azam, M., Pattipati, K., Allanach, J., Poll, S., Patterson-Hine, A., 2005. In-flight fault detection and isolation in aircraft flight control systems. In: 2005 IEEE Aerospace Conference. IEEE, pp. 3555–3565. http://dx.doi.org/10.1109/AERO.2005.1559659.

Bajracharya, M., Maimone, M.W., Helmick, D., 2008. Autonomy for mars rovers: Past, present, and future. Computer 41 (12), http://dx.doi.org/10.1109/MC.2008.479.

Banerjee, T.P., Das, S., 2013. Intelligent fault tracking by an adaptive fuzzy predictor and a fractional controller of electromechanical system–a hybrid approach. In: International Conference on Swarm, Evolutionary, and Memetic Computing. Springer, pp. 574–582. http://dx.doi.org/10.1007/978-3-319-03756-1_51.

Barata, J., Ribeiro, L., Onori, M., 2007. Diagnosis on evolvable production systems. In: 2007 IEEE International Symposium on Industrial Electronics. IEEE, pp. 3221–3226. http://dx.doi.org/10.1109/ISIE.2007.4375131.

Bartocci, E., Ferrère, T., Manjunath, N., Ničković, D., 2018. Localizing faults in simulink/stateflow models with STL. In: Proceedings of the 21st International Conference on Hybrid Systems: Computation and Control (Part of CPS Week). ACM, pp. 197–206. http://dx.doi.org/10.1145/3178126.3178131.

Benowitz, E., 2014. The curiosity mars rover's fault protection engine. In: 2014 IEEE International Conference on Space Mission Challenges for Information Technology. IEEE, pp. 62–66. http://dx.doi.org/10.1109/SMC-IT.2014.16.

von Birgelen, A., Buratti, D., Mager, J., Niggemann, O., 2018. Self-organizing maps for anomaly localization and predictive maintenance in cyber-physical production systems. Procedia CIRP 72, 480–485. http://dx.doi.org/10.1016/j.procir.2018.03.150.

Bolbot, V., Theotokatos, G., Bujorianu, M.L., Boulougouris, E., Vassalos, D., 2018. Vulnerabilities and safety assurance methods in cyber-physical systems: A comprehensive review. Reliab. Eng. Syst. Saf..

Bradatsch, C., Ungerer, T., Zalman, R., Lajtkep, A., 2011. Towards runtime testing in automotive embedded systems. In: Industrial Embedded Systems (SIES), 2011 6th IEEE International Symposium on. IEEE, pp. 55–58. http://dx.doi.org/10.1109/SIES.2011.5953679.

Bregon, A., Daigle, M., Roychoudhury, I., Biswas, G., Koutsoukos, X., Pulido, B., 2014. An event-based distributed diagnosis framework using structural model decomposition. Artificial Intelligence 210, 1–35.

Cabasino, M.P., Seatzu, C., Mahulea, C., Silva, M., 2010. Fault diagnosis of manufacturing systems using continuous Petri nets. In: 2010 IEEE International Conference on Systems, Man and Cybernetics. IEEE, pp. 534–539. http://dx.doi.org/10.1109/ICSMC.2010.5642021.

Cacchione, P.Z., 2016. The Evolving Methodology of Scoping Reviews. SAGE Publications Sage CA, Los Angeles, CA, http://dx.doi.org/10.1177/1054773816637493.

Castaldi, P., Mimmo, N., Simani, S., 2016. Fault diagnosis and fault tolerant control strategies for aerospace systems. In: 2016 3rd Conference on Control and Fault-Tolerant Systems (SysTol). IEEE, pp. 684–689. http://dx.doi.org/10.1109/SYSTOL.2016.7739828.

Castleberry, A., Nolen, A., 2018. Thematic analysis of qualitative research data: Is it as easy as it sounds? Curr. Pharm. Teach. Learn. 10 (6), 807–815. http://dx.doi.org/10.1016/j.cptl.2018.03.019.

Chamseddine, A., Amoozgar, M.H., Zhang, Y.M., 2015. Experimental validation of fault detection and diagnosis for unmanned aerial vehicles. In: Handbook of Unmanned Aerial Vehicles. Springer, pp. 1123–1155. http://dx.doi.org/10.1007/978-90-481-9707-1_41.

Chang, X., Huang, J., Lu, F., 2017. Robust in-flight sensor fault diagnostics for aircraft engine based on sliding mode observers. Sensors 17 (4), 835. http://dx.doi.org/10.3390/s17040835.

Charette, R.N., 2009. This car runs on code. IEEE Spectr. 46 (3), 3.

Chen, Y.-g., 2011. Applications of Bayesian network in fault diagnosis of braking system. In: Intelligent Human-Machine Systems and Cybernetics (IHMSC), 2011 International Conference on, Vol. 1. IEEE, pp. 234–237. http://dx.doi.org/10.1109/IHMSC.2011.63.

Chen, Z., Liu, X., Zhang, R., Liu, H., 2014. An automotive electronic throttle testing equipment based on STM32. In: Computer, Consumer and Control (IS3C), 2014 International Symposium on. IEEE, pp. 478–481. http://dx.doi.org/10.1109/IS3C.2014.131.

Chen, Z., Yang, Y., Hu, Z., 2012. A technical framework and roadmap of embedded diagnostics and prognostics for complex mechanical systems in prognostics and health management systems. IEEE Trans. Reliab. 61 (2), 314–322. http://dx.doi.org/10.1109/TR.2012.2196171.

Chu, J., Zhang, L., Cui, P., 2008. Study on integration diagnosis system for auto-mobile faults and its key technologies. In: 2008 IEEE Pacific-Asia Workshop on Computational Intelligence and Industrial Application, Vol. 1. IEEE, pp. 159–162. http://dx.doi.org/10.1109/PACIIA.2008.207.

Cossé, R., Berdjag, D., Piechowiak, S., Duvivier, D., Gaurel, C., 2015. Meta-diagnosis for a special class of cyber-physical systems: The avionics test benches. In: International Conference on Industrial, Engineering and Other Applications of Applied Intelligent Systems. Springer, pp. 635–644. http://dx.doi.org/10.1007/978-3-319-19066-2_61.

Council, N.R., 2001. Embedded, Everywhere: A Research Agenda for Networked Systems of Embedded Computers. The National Academies Press, http://dx.doi.org/10.17226/10193.

Cruzes, D.S., Dyba, T., 2011. Recommended steps for thematic synthesis in software engineering. In: 2011 International Symposium on Empirical Software Engineering and Measurement. IEEE, pp. 275–284. http://dx.doi.org/10.1109/ESEM.2011.36.

Cruzes, D.S., Dybå, T., 2011. Research synthesis in software engineering: A tertiary study. Inf. Softw. Technol. 53 (5), 440–455. http://dx.doi.org/10.1016/j.infsof.2011.01.004.

De Kleer, J., Williams, B.C., 1987. Diagnosing multiple faults. Artificial Intelligence 32 (1), 97–130. http://dx.doi.org/10.1016/0004-3702(87)90063-4.

Dearden, R., Willeke, T., Simmons, R., Verma, V., Hutter, F., Thrun, S., 2004. Real-time fault detection and situational awareness for rovers: Report on the mars technology program task. In: 2004 IEEE Aerospace Conference Proceedings (IEEE Cat. No. 04TH8720), Vol. 2. IEEE, pp. 826–840. http://dx.doi.org/10.1109/AERO.2004.1367683.

Dowdeswell, B., Sinha, R., MacDonell, S., 2019. Data set finding faults: a scoping study of fault diagnostics for industrial cyber-physical systems. In: Mendeley Data, v1. Elsevier, http://dx.doi.org/10.17632/wg9shy9rsm.1.

Dragojević, M., Stević, S., Stupar, G., Živkov, D., 2018. Utilizing IoT technologies for remote diagnostics of next generation vehicles. In: 2018 IEEE 8th International Conference on Consumer Electronics-Berlin (ICCE-Berlin). IEEE, pp. 1–4. http://dx.doi.org/10.1109/ICCE-Berlin.2018.8576249.

Ducard, G., 2013. The SMAC fault detection and isolation scheme: Discussions, improvements, and application to a UAV. In: 2013 Conference on Control and Fault-Tolerant Systems (SysTol). IEEE, pp. 480–485. http://dx.doi.org/10.1109/SysTol.2013.6693949.

Dwyer, M.B., Purandare, R., Person, S., 2010. Runtime verification in context: Can optimizing error detection improve fault diagnosis? In: International Conference on Runtime Verification. Springer, pp. 36–50. http://dx.doi.org/10.1007/978-3-642-16612-9_4.

EARTO, 2014. The TRL Scale as a Research and Innovation Policy Tool, EARTO Recommendations. European Association of Research & Technology Organisations, [cited 27 July 2019]. URL https://www.earto.eu/wp-content/uploads/The_TRL_Scale_as_a_R_I_Policy_Tool_-_EARTO_Recommendations_-_Final.pdf.

Ebert, C., Favaro, J., 2017. Automotive software. IEEE Softw. (3), 33–39. http://dx.doi.org/10.1109/MS.2017.82.

Efkemann, C., Hartmann, T., 2008. Specification of conditions for error diagnos- tics. Electron. Notes Theor. Comput. Sci. 217, 97–112. http://dx.doi.org/10.1016/j.entcs.2008.06.044.

ESA, 2009. ESA Software Engineering and Standardization. European Space Agency Software and Standards Group, [cited 11 July 2017]. URL http://http://www.esa.int/TEC/Software_engineering_and_standardisation/ TECP5EUXBQE_0.html.

Fang, H., Shi, H., Dong, Y., Fan, H., Ren, S., 2017. Spacecraft power system fault diagnosis based on DNN. In: 2017 Prognostics and System Health Management Conference (PHM-Harbin). IEEE, pp. 1–5. http://dx.doi.org/10.1109/PHM.2017.8079271.

Feiler, P., Gluch, D., Mcgregor, J., 2016. An architecture-led safety analysis method. In: 8th European Congress on Embedded Real Time Software and Systems (ERTS 2016), Toulouse.

Feiler, P.H., Lewis, B.A., Vestal, S., 2006. The SAE architecture analysis & design language (AADL) a standard for engineering performance critical systems. In: 2006 IEEE Conference on Computer Aided Control System Design, 2006 IEEE International Conference on Control Applications, 2006 IEEE International Symposium on Intelligent Control. IEEE, pp. 1206–1211. http://dx.doi.org/10.1109/CACSD-CCA-ISIC.2006.4776814.

Feldman, S.P., 2000. Micro matters: The aesthetics of power in NASA's flight readiness review. J. Appl. Behav. Sci. 36 (4), 474–490. http://dx.doi.org/10.1177/0021886300364005.

Fleischmann, H., Kohl, J., Franke, J., 2016. A modular architecture for the design of condition monitoring processes. Procedia CIRP 57, 410–415. http://dx.doi.org/10.1016/j.procir.2016.11.071.

Folkesson, P., Ayatolahi, F., Sangchoolie, B., Vinter, J., Islam, M., Karlsson, J., 2014. Back-to-back fault injection testing in model-based development. In: International Conference on Computer Safety, Reliability, and Security. Springer, pp. 135–148. http://dx.doi.org/10.1007/978-3-319-24255-2_11.

Fürst, S., Bechter, M., 2016. AUTOSAR for connected and autonomous vehicles: The AUTOSAR adaptive platform. In: 2016 46th Annual IEEE/IFIP International Conference on Dependable Systems and Networks Workshop (DSN-W). IEEE, pp. 215–217. http://dx.doi.org/10.1109/DSN-W.2016.24.

Ghadhab, M., Kuntz, M., Kuvaiskii, D., Fetzer, C., 2015. A controller safety concept based on software-implemented fault tolerance for fail-operational automotive applications. In: International Workshop on Formal Techniques for Safety-Critical Systems. Springer, pp. 189–205. http://dx.doi.org/10.1007/978-3-319-29510-7_11.

Gill, H., 2008. From vision to reality: cyber-physical systems. In: HCSS National Workshop on New Research Directions for High Confidence Transportation CPS: Automotive, Aviation, and Rail.

Golagha, M., Pretschner, A., Fisch, D., Nagy, R., 2017. Reducing failure analysis time: An industrial evaluation. In: 2017 IEEE/ACM 39th International Conference on Software Engineering: Software Engineering in Practice Track (ICSE-SEIP). IEEE, pp. 293–302. http://dx.doi.org/10.1109/ICSE-SEIP.2017.15.

Gonzalez, J.P.N., Castañon, L.E.G., Rabhi, A., El Hajjaji, A., Morales-Menendez, R., 2009. Vehicle fault detection and diagnosis combining AANN and ANFI. IFAC Proc. Vol. 42 (8), 1079–1084. http://dx.doi.org/10.3182/20090630-4-ES-2003.00178.

Goupil, P., Boada-Bauxell, J., Marcos, A., Cortet, E., Kerr, M., Costa, H., 2014. AIRBUS efforts towards advanced real-time fault diagnosis and fault tol- erant control. IFAC Proc. Vol. 47 (3), 3471–3476. http://dx.doi.org/10.3182/20140824-6-ZA-1003.01945.

Grimm, S., Watzke, M., Hubauer, T., Cescolini, F., 2012. Embedded reasoning on programmable logic controllers. In: International Semantic Web Conference. Springer, pp. 66–81. http://dx.doi.org/10.1007/978-3-642-35173-0_5.

Guo, A., Yu, D., Du, H., Hu, Y., Yin, Z., Li, H., 2016. Cyber-physical failure detection system: Survey and implementation. In: Cyber Technology in Automation, Control, and Intelligent Systems (CYBER), 2016 IEEE International Conference on. IEEE, pp. 428–432. http://dx.doi.org/10.1109/CYBER.2016.7574863.

Haghighatkhah, A., Banijamali, A., Pakanen, O.-P., Oivo, M., Kuvaja, P., 2017. Automotive software engineering: A systematic mapping study. J. Syst. Softw. 128, 25–55. http://dx.doi.org/10.1016/j.jss.2017.03.005.

Hale, W.T., Bollas, G.M., 2018. Design of built-in tests for active fault detection and isolation of discrete faults. IEEE Access 6, 50959–50973. http://dx.doi.org/10.1109/ACCESS.2018.2869269.

Harirchi, F., Ozay, N., 2016. Guaranteed model-based fault detection in cyber- physical systems: A model invalidation approach. arXiv, doi:arXiv:1609.05921.

Harirchi, F., Ozay, N., 2018. Guaranteed model-based fault detection in cyber– physical systems: A model invalidation approach. Automatica 93, 476–488. http://dx.doi.org/10.1016/j.automatica.2018.03.040.

Héder, M., 2017. From NASA to EU: The evolution of the TRL scale in public sector innovation. Innov. J. 22 (2), 1–23.

Hegde, R., Mishra, G., Gurumurthy, K., 2011. An insight into the hardware and software complexity of ECUs in vehicles. In: International Conference on Advances in Computing and Information Technology. Springer, pp. 99–106. http://dx.doi.org/10.1007/978-3-642-22555-0_11.

Heirung, T.A.N., Mesbah, A., 2019. Input design for active fault diagnosis. Annu. Rev. Control http://dx.doi.org/10.1016/j.arcontrol.2019.03.002.

Hieb, J., Graham, J., Guan, J., 2009. An ontology for identifying cyber intrusion induced faults in process control systems. In: Critical Infrastructure Protection III. Springer, pp. 125–138. http://dx.doi.org/10.1007/978-3-642-04798-5_9.

Höfig, K., Zeller, M., Schorp, K., 2015. Automated failure propagation using inner port dependency traces. In: Proceedings of the 11th International ACM SIGSOFT Conference on Quality of Software Architectures. ACM, pp. 123–128. http://dx.doi.org/10.1145/2737182.2737191.

Holzmann, G.J., 2014. Mars code. Commun. ACM 57 (2), 64–73. http://dx.doi.org/ 10.1145/2560217.2560218.

Hu, H., Li, Z., Al-Ahmari, A., 2011. Reversed fuzzy Petri nets and their application for fault diagnosis. Comput. Ind. Eng. 60 (4), 505–510. http://dx.doi.org/10. 1016/j.cie.2010.12.003.

Huang, X.-F., Zhou, C.-J., Huang, S., Huang, K.-X., Li, X., 2014. Transient fault detection in networked control systems. Int. J. Distrib. Sens. Netw. 10 (11), 346269. http://dx.doi.org/10.1155/2014/346269.

ISO/BSS, 2019. Space Systems. Definition of the Technology Readiness Levels (TRLs) and their Criteria of Assessment. ISO/BSS.

ISO/IEC, 2000. (E)-Road Vehicles Diagnostic Systems Keyword Protocol 2000. ISO/IEC.

Iverson, D.L., Martin, R., Schwabacher, M., Spirkovska, L., Taylor, W., Mackey, R., Castle, J.P., Baskaran, V., 2012. General purpose data-driven monitoring for space operations. J. Aerosp. Comput. Inf. Commun. 9 (2), 26–44. http://dx.doi.org/10.2514/1.54964.

Jacoby, G., et al., 2010. Testing adaptive probabilistic software components in cyber systems. In: Monterey Workshop. Springer, pp. 228–238. http://dx. doi.org/10.1007/978-3-642-21292-5_13.

Janasak, K.M., Beshears, R.R., 2007. Diagnostics to prognostics- A product avail- ability technology evolution. In: Reliability and Maintainability Symposium, 2007. RAMS'07. Annual. IEEE, pp. 113–118. http://dx.doi.org/10.1109/RAMS.2007.328051.

Jazdi, N., 2014. Cyber physical systems in the context of industry 4.0. In: Au- tomation, Quality and Testing, Robotics, 2014 IEEE International Conference on. IEEE, pp. 1–4. http://dx.doi.org/10.1109/AQTR.2014.6857843.

Jiang, Y., Li, K., Yin, S., 2018. Cyber-physical system based factory monitoring and fault diagnosis framework with plant-wide performance optimization. In: 2018 IEEE Industrial Cyber-Physical Systems (ICPS). IEEE, pp. 240–245. http://dx.doi.org/10.1109/ICPHYS.2018.8387666.

Johnson, D.M., 1996. A review of fault management techniques used in safety- critical avionic systems. Prog. Aerosp. Sci. 32 (5), 415–431. http://dx.doi.org/ 10.1016/0376-0421(96)82785-0.

Kane, A., Fuhrman, T., Koopman, P., 2014. Monitor based oracles for cyber- physical system testing: Practical experience report. In: 2014 44th Annual IEEE/IFIP International Conference on Dependable Systems and Networks. IEEE, pp. 148–155. http://dx.doi.org/10.1109/DSN.2014.28.

Käßmeyer, M., Berndt, R., Bazan, P., German, R., 2016. Product line fault tree analysis by means of multi-valued decision diagrams. In: International GI/ITG Conference on Measurement, Modelling, and Evaluation of Computing Systems and Dependability and Fault Tolerance. Springer, pp. 122–136. http: //dx.doi.org/10.1007/978-3-319-31559-1_11.

Khlif, M., Shawky, M., 2008. Enhancing diagnosis ability for embedded electronic systems using co-modeling. In: Novel Algorithms and Techniques in Telecommunications, Automation and Industrial Electronics. Springer, pp. 143–149.

Khlif, M., Tahan, O., Shawky, M., 2010. Co-simulation trace analysis (COSITA) tool for vehicle electronic architecture diagnosability analysis. In: Intelligent Vehicles Symposium (IV), 2010 IEEE. IEEE, pp. 572–578. http://dx.doi.org/10.1109/IVS.2010.5548045.

Khorasgani, H., Biswas, G., Jung, D., 2019. Structural methodologies for distributed fault detection and isolation. Appl. Sci. 9 (7), 1286. http://dx.doi.org/10.3390/app9071286.

Khoukhi, A., Khalid, M.H., 2015. Hybrid computing techniques for fault detection and isolation, a review. Comput. Electr. Eng. 43, 17–32. http://dx.doi.org/10.1016/j.compeleceng.2014.12.015.

Kim, M.H., Lee, S., Lee, K.C., 2011. A fuzzy predictive redundancy system for fault-tolerance of x-by-wire systems. Microprocess. Microsyst. 35 (5), 453–461. http://dx.doi.org/10.1016/j.micpro.2011.04.003.

Kitchenham, B., Pretorius, R., Budgen, D., Brereton, O.P., Turner, M., Niazi, M., Linkman, S., 2010. Systematic literature reviews in software engineering–a tertiary study. Inf. Softw. Technol. 52 (8), 792–805. http://dx.doi.org/10.1016/ j.infsof.2010.03.006.

Klar, D., Huhn, M., 2012. Interfaces and models for the diagnosis of cyber- physical ecosystems. In: Digital Ecosystems Technologies (DEST), 2012 6th IEEE International Conference on. IEEE, pp. 1–6. http://dx.doi.org/10.1109/DEST.2012.6227948.

Ko, D., Kim, T., Park, J., Kang, S., Chun, I., 2016. An approach to applying goal model and fault tree for autonomic control. Contemp. Eng. Sci. 9, 843–851. http://dx.doi.org/10.12988/ces.2016.6697.

Kodali, A., Zhang, Y., Sankavaram, C., Pattipati, K., Salman, M., 2012. Fault diagnosis in the automotive electric power generation and storage system (EPGS). IEEE/ASME Trans. Mechatronics 18 (6), 1809–1818. http://dx.doi.org/10.1109/TMECH.2012.2214397.

Kodali, A., Zhang, Y., Sankavaram, C., Pattipati, K., Salman, M., 2013. Fault diagnosis in the automotive electric power generation and storage system (EPGS). IEEE/ASME Trans. Mechatronics 18 (6), 1809–1818. http://dx.doi.org/10.1109/TMECH.2012.2214397.

Koitz, R., Lüftenegger, J., Wotawa, F., 2017. Model-based diagnosis in practice: interaction design of an integrated diagnosis application for industrial wind turbines. In: International Conference on Industrial, Engineering and Other Applications of Applied Intelligent Systems. Springer, pp. 440–445. http: //dx.doi.org/10.1007/978-3-319-60042-0_48.

Kunst, N., Judkins, J., Lynn, C., Goodman, D., 2009. Damage propagation analysis methodology for electromechanical actuator prognostics. In: 2009 IEEE Aerospace Conference. IEEE, pp. 1–7. http://dx.doi.org/10.1109/DSN.2014.28.

Kurien, J., Koutsoukos, X., Zhao, F., 2002. Distributed Diagnosis of Networked, Embedded Systems. Tech. rep., XEROX Palo Alto Research Center CA.

Kurz, D., Kaspar, J., Pilz, J., 2011. Dynamic maintenance in semiconductor manufacturing using Bayesian networks. In: Automation Science and Engineering (CASE), 2011 IEEE Conference on. IEEE, pp. 238–243. http://dx.doi.org/10.1109/CASE.2011.6042404.

Langer, F., Eilers, D., Knorr, R., 2009. Fault detection in discrete event based distributed systems by forecasting message sequences with neural networks. In: Annual Conference on Artificial Intelligence. Springer, pp. 411–418. http://dx.doi.org/10.1007/978-3-642-04617-9_52.

Lapira, E.R., Bagheri, B., Zhao, W., Lee, J., Henriques, R.V., Pereira, C.E., Piccoli, L., Guimarães, C., 2013. A systematic approach to intelligent maintenance of production systems with a framework for embedded implementation. IFAC Proc. Vol. 46 (7), 23–28. http://dx.doi.org/10.3182/20130522-3-BR-4036.00092.

Laughton, M.A., Say, M.G., 2013. Electrical Engineer's Reference Book. Elsevier.

Le Mortellec, A., Clarhaut, J., Sallez, Y., Berger, T., Trentesaux, D., 2013. Embedded holonic fault diagnosis of complex transportation systems. Eng. Appl. Artif. Intell. 26 (1), 227–240. http://dx.doi.org/10.1016/j.engappai.2012.09.008.

Lee, E.A., 2006. Cyber-physical systems-are computing foundations adequate. In: Position Paper for NSF Workshop on Cyber-Physical Systems: Research Motivation, Techniques and Roadmap, Vol. 2. Citeseer, pp. 1–9.

Lee, J., Bagheri, B., Kao, H.-A., 2015. A cyber-physical systems architecture for industry 4.0-based manufacturing systems. Manuf. Lett. 3, 18–23. http://dx.doi.org/10.1016/j.mfglet.2014.12.001.

Lee, J., Jin, C., Bagheri, B., 2017. Cyber physical systems for predictive production systems. Prod. Eng. 11 (2), 155–165.

Lee, E.A., Seshia, S.A., 2016. Introduction to Embedded Systems: A Cyber-Physical Systems Approach. MIT Press.

Leitão, P., Karnouskos, S., Ribeiro, L., Moutis, P., Barbosa, J., Strasser, T.I., 2017. Common practices for integrating industrial agents and low level automation functions. In: IECON 2017-43rd Annual Conference of the IEEE Industrial Electronics Society. IEEE, pp. 6665–6670. http://dx.doi.org/10.1109/IECON.2017.8217164.

Levac, D., Colquhoun, H., O'Brien, K.K., 2010. Scoping studies: advancing the methodology. Implement. Sci. 5 (1), 69. http://dx.doi.org/10.1186/1748-5908-5-69.

Lira, D.N., Borsato, M., 2016. Dependability modeling for the failure prognostics in smart manufacturing. In: ISPE TE. pp. 885–894. http://dx.doi.org/10.3233/978-1-61499-703-0-885.

Losik, L., 2009. Using Generic Telemetry Prognostic Algorithms for Launch Vehicle and Spacecraft Independent Failure Analysis Service. International Foundation for Telemetering.

Losik, L., 2011. Using Data-Driven Prognostic Algorithms for Completing Independent Failure Analysis. International Foundation for Telemetering. Lu, Y., Dong, Y., Wei, X., Xiao, M., 2018. A hybrid method of redundancy system reliability analysis based on AADL models. In: 2018 IEEE International Conference on Software Quality, Reliability and Security Companion (QRS-C). IEEE, pp. 294–300. http://dx.doi.org/10.1109/QRS-C.2018.00060.

Luo, J., Choi, K., Pattipati, K.R., Qiao, L., Chigusa, S., 2006. Distributed fault diagnosis for networked, embedded automotive systems. In: 2006 IEEE International Conference on Systems, Man and Cybernetics, Vol. 2. IEEE, pp. 1226–1232. http://dx.doi.org/10.1109/ICSMC.2006.384882.

Mankins, J.C., 1995. Technology Readiness Levels. White Paper.

Mankins, J.C., 2009. Technology readiness assessments: A retrospective. Acta Astronaut. 65 (9–10), 1216–1223. http://dx.doi.org/10.1016/j.actaastro.2009.03.058.

Marzat, J., Piet-Lahanier, H., Damongeot, F., Walter, E., 2009. A new model-free method performing closed-loop fault diagnosis for an aeronautical system. In: 7th Workshop on Advanced Control and Diagnosis, ACD'2009. p. 6. http://dx.doi.org/10.3182/20090706-3-FR-2004.00032.

Mays, N., Roberts, E., Popay, J., et al., 2001. Synthesising research evidence. In: Studying the Organisation and Delivery of Health Services: Research Methods, Vol. 220.

McGregor, J.D., Gluch, D.P., Feiler, P.H., 2017. Analysis and design of safety-critical, cyber-physical systems. ACM SIGAda Ada Lett. 36 (2), 31–38. Milis, G.M., Eliades, D.G., Panayiotou, C.G., Polycarpou, M.M., 2016. A cognitive fault-detection design architecture. In: Neural Networks (IJCNN), 2016 International Joint Conference on. IEEE, pp. 2819–2826. http://dx.doi.org/10.1109/IJCNN.2016.7727555.

MIMOSA, 2019. Machinery Information Management Open Systems Alliance (MIMOSA). [cited 29 June, 2019]. URL https://www.mimosa.org.

Modest, C., Thielecke, F., 2016. SPYDER: a software package for system diagnosis engineering. CEAS Aeronaut. J. 7 (2), 315–331. http://dx.doi.org/10.1007/s13272-016-0189-0.

Mohrle, F., Zeller, M., Hofig, K., Rothfelder, M., Liggesmeyer, P., 2015. Automated compositional safety analysis using component fault trees. In: Software Reliability Engineering Workshops (ISSREW), 2015 IEEE International Symposium on. IEEE, pp.152–159. http://dx.doi.org/10.1109/ISSREW.2015.7392061.

Munn, Z., Peters, M.D., Stern, C., Tufanaru, C., McArthur, A., Aromataris, E., 2018. Systematic review or scoping review? Guidance for authors when choosing between a systematic or scoping review approach. BMC Med. Res. Methodol. 18 (1), 143. http://dx.doi.org/10.1186/s12874-018-0611-x.

Nagorny, K., Scholze, S., Barata, J., Colombo, A.W., 2016. An approach for implementing ISA 95-compliant big data observation, analysis and diagnosis features in industry 4.0 vision following manufacturing systems. In: Doctoral Conference on Computing, Electrical and Industrial Systems. Springer, pp. 116–123. http://dx.doi.org/10.1007/978-3-319-31165-4_12.

Nasri, O., Lakhal, N.M.B., Adouane, L., Slama, J.B.H., 2019. Automotive decentralized diagnosis based on CAN real-time analysis. J. Syst. Archit. http://dx.doi.org/10.1016/j.sysarc.2019.01.009.

de Normalisation, O.I., 2011. Systems and Software Engineering: Systems and Software Quality Requirements and Evaluation (SQuaRE): System and Software Quality Models. ISO/IEC.

Nuñez, D.L., Borsato, M., 2017. An ontology-based model for prognostics and health management of machines. J. Ind. Inf. Integr. 6, 33–46. http://dx.doi.org/10.1016/j.jii.2017.02.006.

OMG, 2020. What's Driving the Connected Car? OMG, [cited 31 March 2020]. URL https://www.omg.org/omgmarte/.

Pakala, H.G.M., Raju, K., Khan, I., 2011. Integration testing of multiple embedded processing components. In: International Conference on Computer Science and Information Technology. Springer, pp. 200–209. http://dx.doi.org/10.1007/978-3-642-17881-8_20.

Palladino, A., Fiengo, G., Lanzo, D., 2012. A portable hardware-in-the-loop (HIL) device for automotive diagnostic control systems. ISA Trans. 51 (1), 229–236. http://dx.doi.org/10.1016/j.isatra.2011.10.009.

Parr, E.A., 1998. Industrial Control Handbook. Industrial Press Inc..

Pons, R., Subias, A., Travé-Massuyès, L., 2015. Iterative hybrid causal model based diagnosis: Application to automotive embedded functions. Eng. Appl. Artif. Intell. 37, 319–335. http://dx.doi.org/10.1016/j.engappai.2014.09.016.

Procter, S., Feiler, P., 2020. The AADL error library: An operationalized taxonomy of system errors. ACM SIGAda Ada Lett. 39 (1), 63–70. http://dx.doi.org/10.1145/3379106.3379113.

Provan, G., 2014. A contracts-based framework for systems modeling and embedded diagnostics. In: International Conference on Software Engineering and Formal Methods. Springer, pp. 131–143. http://dx.doi.org/10.1007/978-3-319-15201-1_9.

Ramos, A.V., Delamer, I.M., Lastra, J.L., 2011. Embedded service oriented monitoring, diagnostics and control: Towards the asset-aware and self-recovery factory. In: Industrial Informatics (INDIN), 2011 9th IEEE International Conference on. IEEE, pp. 497–502. http://dx.doi.org/10.1109/INDIN.2011.6034930.

Ribeiro, L., Barata, J., 2011. Re-thinking diagnosis for future automation systems: An analysis of current diagnostic practices and their applicability in emerging IT based production paradigms. Comput. Ind. 62 (7), 639–659. http://dx.doi.org/10.1016/j.compind.2011.03.001.

Rizzoni, G., Onori, S., Rubagotti, M., 2009. Diagnosis and prognosis of automotive systems: motivations, history and some results. IFAC Proc. Vol. 42 (8), 191–202. http://dx.doi.org/10.3182/20090630-4-ES-2003.00032.

Ruppert, T., Abonyi, J., 2018. Software sensor for activity-time monitoring and fault detection in production lines. Sensors 18 (7), 2346. http://dx.doi.org/10.3390/s18072346.

Sanislav, T., Mois, G., Miclea, L., 2015. A new approach towards increasing cyber-physical systems dependability. In: Proceedings of the 2015 16th International Carpathian Control Conference (ICCC). IEEE, pp. 443–447. http://dx.doi.org/10.1109/CarpathianCC.2015.7145120.

Sankavaram, C., Kodali, A., Pattipati, K., 2013. An integrated health management process for automotive cyber-physical systems. In: 2013 International Conference on Computing, Networking and Communications (ICNC). IEEE, pp. 82–86. http://dx.doi.org/10.1109/ICCNC.2013.6504058.

Sankavaram, C., Kodali, A., Pattipati, K., Singh, S., Zhang, Y., Salman, M., 2016. An inference-based prognostic framework for health management of automotive systems. Int. J. Progn. Health Manage. 7, 1–16.

Saracco, R., 2016. Guess what requires 150 million lines of code. EIT Digit. Internet Available: https://www.eitdigital.eu/news-events/blog/article/guess-what-requires-150-million-lines-of-code/ (13 Jan. 2016).

Sargolzaei, A., Crane, C.D., Abbaspour, A., Noei, S., 2016. A machine learning approach for fault detection in vehicular cyber-physical systems. In: Machine Learning and Applications (ICMLA), 2016 15th IEEE International Conference on. IEEE, pp. 636–640.

Schoeller, M., Roemer, M., Leonard, M., Derriso, M., 2007. Embedded reasoning supporting aerospace IVHM. In: AIAA Infotech@ Aerospace 2007 Conference and Exhibit. p. 2820. http://dx.doi.org/10.2514/6.2007-2820.

Schubert, U., Kruger, U., Arellano-Garcia, H., de Sá Feital, T., Wozny, G., 2011. Unified model-based fault diagnosis for three industrial application studies. Control Eng. Pract. 19 (5), 479–490. http://dx.doi.org/10.1016/j.conengprac.2011.01.009.

Schulte, P.Z., 2018. A State Machine Architecture for Aerospace Vehicle Fault Protection (Ph.D. thesis). Georgia Institute of Technology.

Schwabacher, M., Aguilar, R., Figueroa, F., 2009. Using decision trees to detect and isolate simulated leaks in the J-2X rocket engine. In: 2009 IEEE Aerospace Conference. IEEE, pp. 1–7. http://dx.doi.org/10.1109/AERO.2009.4839691.

Schwabacher, M., Goebel, K., 2007. A survey of artificial intelligence for prognostics. In: Aaai Fall Symposium. pp. 107–114.

Schwabacher, M., Martin, R., Waterman, R., Oostdyk, R., Ossenfort, J., Matthews, B., 2010. Ares IX ground diagnostic prototype. In: AIAA Infotech@ Aerospace 2010. p. 3354. http://dx.doi.org/10.2514/6.2010-3354.

Schwabacher, M., Waterman, R., 2008. Pre-launch diagnostics for launch vehicles. In: 2008 IEEE Aerospace Conference. IEEE, pp. 1–8.

Schweppe, H., Zimmermann, A., Grill, D., 2009. Flexible on-board stream processing for automotive sensor data. IEEE Trans. Ind. Inf. 6 (1), 81–92. http://dx.doi.org/10.1109/TII.2009.2037145.

Shraim, H., Awada, A., Youness, R., 2018. A survey on quadrotors: Configurations, modeling and identification, control, collision avoidance, fault diagnosis and tolerant control. IEEE Aerosp. Electron. Syst. Mag. 33 (7), 14–33. http://dx.doi.org/10.1109/MAES.2018.160246.

Song, F., Hou, W., Shi, L., 2013. The information-enhanced fault diagnosis system design of avionics power supply module. In: Quality, Reliability, Risk, Maintenance, and Safety Engineering (QR2MSE), 2013 International Conference on. IEEE, pp. 1758–1761. http://dx.doi.org/10.1109/QR2MSE.2013.6625916.

Song, Y.H., Kim, M.H., Lee, S., Lee, K.C., 2010. Implementation of a fuzzy predictive redundancy system for tolerance of x-by-wire systems. In: IECON 2010-36th Annual Conference on IEEE Industrial Electronics Society. IEEE, pp. 3141–3145. http://dx.doi.org/10.1109/IECON.2010.5675027.

Starek, J.A., Açıkmeşe, B., Nesnas, I.A., Pavone, M., 2016. Spacecraft autonomy challenges for next-generation space missions. In: Advances in Control System Technology for Aerospace Applications. Springer, pp. 1–48. http://dx.doi.org/10.1007/978-3-662-47694-9_1.

Steinkamp, N., Levine, R., Roth, R., 2019. 2019 Automotive Defect and Recall Report. Stout Risius Ross, LLC, [cited 18 October, 2019]. URL https://www.stout.com/en/insights/report/2019-automotive-defect-and-recall-report.

Straub, J., 2015. In search of technology readiness level (TRL) 10. Aerosp. Sci. Technol. 46, 312–320.

Struss, P., 2013. Model-based analysis of embedded systems: Placing it upon its feet instead of on its head-an outsider's view. In: ICSOFT. pp. 284–291. http://dx.doi.org/10.5220/0004596102840291.

Sunny, S.M.N.A., Liu, X., Shahriar, M.R., 2018. Remote monitoring and online testing of machine tools for fault diagnosis and maintenance using mtcomm in a cyber-physical manufacturing cloud. In: 2018 IEEE 11th International Conference on Cloud Computing (CLOUD). pp. 532–539. http://dx.doi.org/10.1109/CLOUD.2018.00074.

Swearingen, K., Keller, K., 2007. Health ready systems. In: Autotestcon, 2007 IEEE. IEEE, pp. 625–631. http://dx.doi.org/10.1109/AUTEST.2007.4374277.

Syed, W.A., Khan, S., Phillips, P., Perinpanayagam, S., 2013. Intermittent fault finding strategies. Procedia CIRP 11, 74–79. http://dx.doi.org/10.1016/j.procir.2013.07.062.

Tanveer, A., Sinha, R., MacDonell, S., 2018. On Design-time Security in IEC 61499 Systems: Conceptualisation, Implementation, and Feasibility. IEEE, http://dx.doi.org/10.1109/INDIN.2018.8472093.

Technavio, 2018. Global Commercial Vehicle Remote Diagnostics Market 2018-2022. Technavio, [cited 27 July 2019]. URL https://www.technavio.com.

Terrile, R.J., Doumani, F.G., Ho, G.Y., Jackson, B.L., 2015. Calibrating the technology readiness level (TRL) scale using NASA mission data. In: 2015 IEEE Aerospace Conference. IEEE, pp. 1–9. http://dx.doi.org/10.1109/AERO.2015.7119313.

Thombare, T.R., Dole, L., 2014. Review on fault diagnosis model in automobile. In: Computational Intelligence and Computing Research (ICCIC), 2014 IEEE International Conference on. IEEE, pp. 1–4. http://dx.doi.org/10.1109/ICCIC.2014.7238546.

Trawczynski, D., Sosnowski, J., Gawkowski, P., 2008. Analyzing fault susceptibility of ABS microcontroller. Comput. Saf. Reliab. Secur. 360–372. http://dx.doi.org/10.1007/978-3-540-87698-4_30.

Vong, C.-M., Wong, P.-K., Ip, W.-F., 2011. Simultaneous faults diagnosis for automotive ignition patterns. In: 2011 International Conference on Machine Learning and Cybernetics, Vol. 3. IEEE, pp. 1324–1330. http://dx.doi.org/10.1109/ICMLC.2011.6016890.

Wang, J., Ye, L., Gao, R.X., Li, C., Zhang, L., 2018. Digital twin for rotating machinery fault diagnosis in smart manufacturing. Int. J. Prod. Res. 1–15. http://dx.doi.org/10.1080/00207543.2018.1552032.

Wang, J., Zhang, L., Duan, L., Gao, R.X., 2017. A new paradigm of cloud-based predictive maintenance for intelligent manufacturing. J. Intell. Manuf. 28 (5), 1125–1137. http://dx.doi.org/10.1007/s10845-015-1066-0.

Waszecki, P., Lukasiewycz, M., Chakraborty, S., 2015. Decentralized diagnosis of permanent faults in automotive E/E architectures. In: Embedded Computer Systems: Architectures, Modeling, and Simulation (SAMOS), 2015 International Conference on. IEEE, pp. 189–196. http://dx.doi.org/10.1109/SAMOS.2015.7363675.

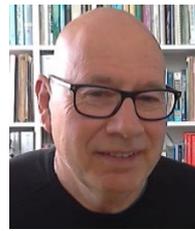

Barry Dowdeswell is a Ph.D. Candidate in The School of Engineering, Computer and Mathematical Sciences at the Auckland University of Technology, New Zealand. His research focuses on fault diagnosis methodologies for industrial-scale cyber–physical systems. His cur- rent work includes the development semi-autonomous, multi-agent based fault diagnostic engines for IoT sys- tems, particularly those built with IEC 61499 Function Blocks.

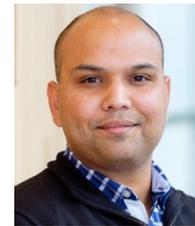

Roopak Sinha (Ph.D. '09, MCE '16, BE (Hons) '03, SFHEA '17) is an Associate Professor in The School of Engineering, Computer and Mathematical Sciences at The Auckland University of Technology, New Zealand. He has previously held academic positions at The University of Auckland, New Zealand and INRIA, France. His research interest is ''Systematic, Standards-First De- sign of Complex, Next-Generation Embedded Software'' applied to domains like Internet-of-Things, Edge Computing, Cyber–Physical Systems, Home and Industrial Automation, and Intelligent Transportation Systems. He has served on several international standardization projects, and works with several New Zealand companies to systematically reduce standards-compliance costs in IoT/embedded products.

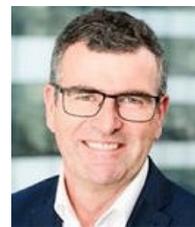

Stephen G. MacDonell is Professor of Software Engineering at Auckland University of Technology and Professor in Information Science at the University of Otago, both in New Zealand. Stephen was awarded BCom(Hons) and MCom degrees from the University of Otago and a Ph.D. from the University of Cambridge. His research has been published in IEEE Transactions on Software Engineering, ACM Transactions on Soft- ware Engineering and Methodology, ACM Computing Surveys, Empirical Software Engineering, Information & Management, the Journal of Systems and Software, Information and Software Technology, and the Project Management Journal, and his research findings have been presented at more than 100 international conferences. He is a Fellow of IT Professionals NZ, Senior Member of the IEEE and the IEEE Computer Society, and Member of the ACM, and he serves on the Editorial Board of Information and Software Technology. Stephen is also Theme Leader for Data Science & Digital Technologies in New Zealand's National Science Challenge Science for Technological Innovation, Technical Advisor to the Office of the Federation of Māori Authorities Pou Whakatāmore Hangarau — Chief Advisor Innovation & Research, and Deputy Chair of Software Innovation New Zealand (SINZ).